\documentclass[preprint]{elsarticle}
\usepackage[english]{babel}
\usepackage{hyperref}
\usepackage[position=b]{subcaption}
\usepackage{tikz} 
\usepackage{pgfplots} 
\pgfplotsset{width=7cm, compat=1.6} 
\usepackage{transparent} 
\usepackage{amsmath}

 \usepackage{geometry}%
\journal{Journal of Computational Physics}
\graphicspath{{pictures/}}

\begin{document}

\begin{frontmatter}

\title{New Approaches for the Interface Reconstruction and Surface Force Computation for Volume of Fluid Simulations of Droplet Interaction of Immiscible Liquids}

\author[mymainaddress]{Johanna Potyka \href{https://orcid.org/0000-0003-1310-4434}{\includegraphics[height=10pt]{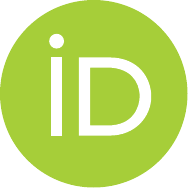}}\corref{mycorrespondingauthor}}
\cortext[mycorrespondingauthor]{Corresponding author}
\ead{johanna.potyka@itlr.uni-stuttgart.de}
\author[mymainaddress]{Kathrin Schulte\href{https://orcid.org/0000-0001-8650-5840}{\includegraphics[height=10pt]{orcid_logo}}}
\address[mymainaddress]{Institut of Aerospace Thermodynamics (ITLR), Pfaffenwaldring 31, 70569 Stuttgart}

\begin{abstract}
The complexity of binary droplet collisions increases for the collision of immiscible liquids with the occurrence of triple lines and thin encapsulating films. The Volume of Fluid (VOF) method is extended with an efficient interface reconstruction applicable to three-component cells of arbitrary configuration. Together with an enhanced Continuous Surface Stress (CSS) model for accurate surface force computation, which was gained by the introduction of a film stabilisation approach, this enables the simulation of the interaction of two droplets of immiscible liquids. With the new methods, simulations of binary droplet collisions of fully wetting liquids were performed with excellent agreement to experimental data in different collision regimes regarding both the morphology and the prediction of the regime boundaries for head-on as well as off-centre collisions.%
\end{abstract}

\begin{keyword}
immiscible liquids \sep droplet collision \sep encapsulation \sep VOF \sep PLIC \sep film stabilization \sep Continuous Surface Stress model \sep contact angle  \sep surface tension
\end{keyword}

\end{frontmatter}


\section{Introduction} \label{Sec:Introduction}
Binary droplet collisions occur in multiple processes in nature as well as in industrial applications. In some applications, such as the injection of water into a diesel engine, the droplets consist of different immiscible liquids \cite{TSURU2010}. The collision of this fully wetting liquid combination of water and diesel was studied in experimental investigations on the binary collision process by Chen~and~Chen~\cite{CHEN2006}. More extensive experimental studies by Planchette~et~al. \cite{PLANCHETTE2009, PLANCHETTE2010, PLANCHETTE2011, PLANCHETTE2012} on fully wetting liquids were dedicated to droplet collisions of immiscible liquids with a focus on micro-encapsulation which is required in pharmaceutical and life science applications as well as food production. Roisman~et.~al.~\cite{ROISMANN2012} provide analytical modelling approaches for the evolution of major morphological parameters such as the collision disc's diameter. Planchette~et.~al.~\cite{PLANCHETTE2017} focus on the analytical modelling of the outcome of head-on binary collisions of immiscible liquids modelled as colliding liquid springs. There is no experimental data available known to the authors of the present study on the droplet collision of partially wetting liquids except a small coalescence study of Wang~et~al.~\cite{WANG2004} which focussed on the burning characteristics of coalesced droplets of Hexadecane and water. Experimental investigations of the influences on the collision morphology and outcome of collisions of fully and partially wetting immiscible liquids are difficult, as the possibilities to vary different parameters and the choice of liquids which are easy to handle in the laboratory is limited. However, there are also only a few numerical studies on the collision of immiscible liquids available in literature. Li~et~al.~\cite{LI2015} obtain a good agreement for the simulation of head-on collisions using a moment of fluid method and an adaptive mesh refinement close to the contact line. They reduce the problem to two dimensions exploiting an axisymmetric assumption of the collision complex. They could capture the regime transition between coalescence and reflexive separation for the head-on collisions presented by Chen and Chen~\cite{CHEN2006}. However, the values for liquid-liquid interfacial tension are not given in either study.
Another numerical study on binary head-on collisions of immiscible droplets was presented recently by Zhang~et~al.~\cite{ZHANG2020}. Their study was motivated by applications in drug delivery and bioprinting. A ternary-fluid diffuse-interface model is applied, again with an adaptive mesh refinement at the liquid-liquid interface. The simulation method is also limited to head-on collisions on a two dimensional grid. They show a good agreement with the experimental results of Planchette~et~al.~\cite{PLANCHETTE2012} for one case of a head-on crossing separation for a Glycerol solution and Silicon Oil M5, a liquid combination also utilised in the present study. \\
The literature shows, that head-on collisions of immiscible liquids are captured well with existing methods by employing a mesh refinement. Nevertheless, the development of a numerical method for asymmetric and three dimensional binary droplet collisions is essential to provide detailed information of this binary collision process, e.g. on the local three dimensional flow field and the contact line movement, and enhance the understanding of the underlying physics. The possibility to vary parameters inside the simulation easily supports the prediction of the outcome of droplet collisions of immiscible liquids. \\
Three dimensional simulations require additional modelling of the interface reconstruction and the surface forces without a computationally expensive adaptive grid refinement. Benson~\cite{BENSON2002} formulated a great need for methods which enable a topology capturing multi-material reconstruction beyond the layered ``onion skin" model to reproduce thin films as well as junctions of the interfaces at the contact line without prescribing the desired configuration. 
In the present work, efficient methods are presented which are able to capture immiscible droplet collisions with any impact parameter and wetting behaviour in three-dimensional simulations. Thus, the presence of multiple interfaces between the three components, the surrounding gas and two immiscible liquids, in arbitrary arrangement inside the same cell of the computational domain can be handled.
The developed methods are implemented into the ITLR's in-house framework for Direct Numerical Simulation of multi-phase flows, Free Surface 3D (FS3D)~\cite{EISENSCHMIDT2016}, to show the application of the presented methods. FS3D employs the well known Volume of Fluid (VOF) approach by Hirt and Nichols~\cite{HIRT1981}. The basis for maintaining a sharp interface between the two phases is a reconstruction of the interface by the Piecewise Linear Interface Calculation (PLIC) \cite{YOUNGS1982, RIDER1998}. The advection of those interfaces is split into three steps by treating the velocity components separately according to the algorithm by Strang et. al.~\cite{STRANG1968}.\\
The here presented reconstruction method for the three interfaces with a formation of a contact line present in the interaction of immiscible liquids in air results from a combination of approaches for two- and three-phase flows existing in literature. Those approaches, which inspire the reconstruction algorithm in this work, are discussed in the following.\\
The idea of PLIC was adapted for two interacting disperse components in a gaseous environment by Pathak and Raessi~\cite{PATHAK2016} as well as by Patel~et~al.~\cite{PATEL2017} who utilise the interface normal's calculation from Washino~et~al.~\cite{WASHINO2010}. Both propose reconstruction algorithms for three-phase problems. The ideas are transferred to immiscible liquids in the present study. Both mentioned approaches are different in their concept and have their own advantages and drawbacks. Pathak and Raessi's reconstruction algorithm is able to reconstruct a great variety of situations without assuming a topology, but as an iterative method it is slow compared to an explicit method. The explicit method proposed by Patel~et~al.~\cite{PATEL2017} on the other hand has the advantage of a fast reconstruction, but does not allow the reconstruction of thin films by always assuming a contact line in a three-component cell.\\
One other idea the present study builds upon and which was originally applied for two-phase flow is the stabilisation of a thin film, which comes with the drawback of not allowing a contact angle if applied in three-phase situations. The general idea of stabilising a lamella or film to avoid unphysical rupturing in the simulations was introduced for symmetrical collisions by Focke and Bothe~\cite{FOCKE2011}. They introduced a stabilising boundary condition in a setup with a symmetry plane thus applicable for head-on collisions.\\
The present work presents methods which efficiently combine the ideas of a film stabilisation with the aforementioned three-phase PLIC algorithms to a topology capturing reconstruction of immiscible liquids.\\
Besides the reconstruction and advection of the interfaces, the surface forces close to the liquid-liquid-gas contact line require additional modelling. The Continuous Surface Force (CSF) model by Brackbill~et~al.~\cite{BRACKBILL1992} was adapted for three-phase flows in numerous studies \cite{PATEL2017, GOEHL2018, WASHINO2010, ANNAPRAGADA2012}. However, the Continuous Surface Stress (CSS) model by Lafaurie~et~al.~\cite{LAFAURIE1994}, which is suitable for large topology changes like separation processes occurring in binary collisions, is employed in this study. For the simulation of the collision of immiscible droplets, this model is adapted in order to account for the two additional interfaces with different interfacial tensions close to the contact line as well as the thin encapsulating film which advances across the inner droplet.\\
In the following, the basic methods implemented in the multi-phase simulation framework FS3D with a focus on pure hydrodynamics are presented in Sec.~\ref{Sec:Basics} as a baseline for this study. Subsequently, the new methods developed to simulate immiscible droplet collisions are explained. A discussion on the challenges arising in VOF simulations of immiscible liquids with thin encapsulating films is summarised in Sec.~\ref{Sec:Problem}. A topology capturing film stabilisation PLIC method as well as modifications for the advection are shown in Sec.~\ref{Sec:3vof}. The computation of the surface forces with an adapted CSS model are presented in Sec.~\ref{Sec:CSS3ph} for two immiscible liquids in a continuous gas phase. Validation results from the implementation in FS3D are presented in Sec.~\ref{Sec:Results}.

\section{Basic Methods for Two-phase Flow} %
\label{Sec:Basics} %
\subsection{Governing Equations} \label{Sec:BasicsEquations} %
The utilised software package FS3D solves the incompressible Navier-Stokes equations to simulate the evolution of multi-phase flows over time $t$. 
Therefore, not only the mass conservation

\begin{equation} 
  \frac{\partial \rho}{\partial t} + \mathbf{u} \cdot \mathbf{\nabla} \rho = 0
\end{equation} 
but also the volume conservation

\begin{equation}
  \mathbf{\nabla} \cdot \mathbf{u} = 0 \label{Eq:Volumeconservation}
\end{equation} %
is fulfilled, providing a divergence free condition for the velocity field $\mathbf{u}$. The momentum conservation

\begin{equation}
  \rho \left[\frac{\partial\mathbf{u}}{\partial t} + \mathbf{\nabla}\cdot \left(\mathbf{u}\mathbf{u}\right)\right] = -\mathbf{\nabla} p + \mathbf{\nabla} \cdot \mathbf{S} +\rho \mathbf{g} + \mathbf{f}_{\Gamma} \label{Eq:Momentumconservation} %
\end{equation} %
is solved with an in this study constant density $\rho$ and the assumption of a Newtonian fluid with the constant dynamic viscosity $\mu$, thus 

\begin{equation}
  \mathbf{S} = \mu \left[\mathbf{\nabla} \mathbf{u}+\left(\mathbf{\nabla} \mathbf{u}\right)^T\right]\text{.}
\end{equation} %
The surface forces $\mathbf{f}_{\Gamma}$ only act in the vicinity of an interface. As the pressure cannot be expressed by the other conserved variables for incompressible flows, the elliptical Pressure-Poisson-equation is solved to obtain the pressure gradient $\mathbf{\nabla} p$. The acceleration due to gravity $\mathbf{g}$ was neglected for this study as the influence is considered small in the cases studied in Sec.~\ref{Sec:Results}. The energy equation is not considered either in this study as isothermal conditions and no phase change are assumed. The modelling of the interfacial forces $\mathbf{f}_\Gamma$ due to the surface tension $\sigma_{l_1-g}$ is thoroughly discussed in the following as new modelling approaches of the interfacial forces as well as the interface reconstruction are essential for the successful simulation of droplet collisions of immiscible liquids.
\subsection{Volume of Fluid Method for Two Components}\label{Sec:VOF} %
Multiple methods exist to numerically solve the Navier-Stokes-Equations for multi-phase flows. The one-field VOF method by Hirt and Nichols \cite{HIRT1981} introduces an indicator function for the presence of the two phases. A non-moving Cartesian and equidistant Marker and Cell grid \cite{HARLOW1965} is utilised in this study to avoid decoupling of the momentum control volumes. In a Cartesian grid the liquid-gas interface does usually not reside exactly at a cell face. Thus, the volume fraction

\begin{equation}
    f_{l_1} = \frac{V_{l_1}}{V_{cell}} =
      \begin{cases}  
         1$, liquid~1 only$ \\ 
         \left]0,1\right[$, at an interface$ \\ 
         0 $, liquid~1 absent$ \\
      \end{cases}      
      \label{Eq:VOFVar}
\end{equation}
accounts for partly filled cells in the discretised problem, with the cell volume $V_{cell}$ of the cuboid cell and the liquid~1's volume $V_{l_1}$ inside that cell. The density and dynamic viscosity

\begin{equation}
\begin{split}
  \rho_{cell} = f_{l_1} \rho_{l_1} + \left(1-f_{l_1}\right) \rho_{gas} \\
  \mu_{cell} = f_{l_1} \mu_{l_1} + \left(1-f_{l_1}\right) \mu_{gas}
  \end{split}
  \label{Eq:Properties}
\end{equation}
are volume weighted averages inside each cell.\\
The volume fraction's advection of liquid~1 (index $l_1$) is expressed as

\begin{equation}
\frac{\partial f_{l_1}}{\partial t} + \mathbf{\nabla} \cdot \left(f_{l_1}\mathbf{u}\right) = 0
\label{Eq:fadvect}
\end{equation}
with the cell average velocity $\mathbf{u}$ from the previous solution of the momentum equation. 
\paragraph{Reconstruction of two phases} As discussed in the introduction, a reconstruction of the interface inside each cell preserves a sharp interface during the subsequent geometric advection of the volume fraction according to Eq.~(\ref{Eq:fadvect}). The Piecewise Linear Interface Calculation (PLIC) by Youngs \cite{YOUNGS1982} and Rider and Kothe \cite{RIDER1998} approximates the interface as a plane in each cell

\begin{equation}
\mathbf{n}_{l_1} \cdot \mathbf{x} = l_{l_1}
\end{equation}
with the interface's normal $\mathbf{n}_{l_1}$ and the distance $l_{l_1}$ to the origin defining the points $\mathbf{x}$ which are on the plane. The interface's normal is the gradient of liquid's volume fraction

\begin{equation}
\mathbf{n}_{l_1} = \mathbf{\nabla} f_{l_1} \label{Eq:GradientNormal}
\end{equation}
which is approximated with central differences. The position $l$ inside the cell is selected such, that the volume corresponding to the volume fraction $f_{l_1}$ is enclosed. The relation between the volume fraction and the position $l$ can be formulated explicitly for the few situations possible for one interface inside a cuboid cell. Thus, the two-phase PLIC method is computationally inexpensive with a Cartesian grid. With the reconstructed planes in each cell, an advection step is performed subsequently.
\paragraph{Advection of Two Phases} The advection of the volume fraction with the velocity $\mathbf{u} = \left( u, v, w\right)^T$ is simplified by splitting it into three subsequent steps. This split advection by Strang \cite{STRANG1968} predicts the volume fluxes across the cell's face by intersecting the reconstructed plane in the adjacent cell at a distance $u \Delta t$ in $x$-direction, $v \Delta t$ in $y$-direction and $w \Delta t$ in $z$-direction from the cell face in interface cells. The order of the three advection steps is permuted every time step $\Delta t$. 
\subsection{Surface Force Modelling for Two Phases}\label{Sec:CSS}
The interfacial forces $\mathbf{f}_\Gamma$ resulting from the surface tension $\sigma_{l_1-g}$ are approximated with the Continuous Surface Stress model (CSS) \cite{LAFAURIE1994} which is chosen for this work as it is suitable for the simulation of the disintegration of the droplets after a collision at higher velocities. The surface forces

\begin{equation}
\mathbf{f}_\Gamma = \mathbf{\nabla} \cdot \left(\sigma_{l_1-g} \tilde{a}_{\Gamma,l_1} \left(\mathbf{I}-\mathbf{\tilde{n}}_{\Gamma,l_1} \otimes \mathbf{\tilde{n}}_{\Gamma,l_1}\right)\right)\label{Eq:fgamma}
\end{equation}
are derived from the divergence of the anisotropic interface stress tensor originating from a mesoscopic view on the pressure jump across the droplet's surface. To represent the mesoscopic interface as a transition zone, the volume fraction field is smoothed. It is reported that smoothing the $f_{l_1}$ volume fraction previous to the calculation of the interface's normals $\mathbf{\tilde{n}}_{l_1}$ with Eq.~\ref{Eq:GradientNormal} applied to the smoothed field leads to a more accurate surface force computation \cite{LAFAURIE1994}. This smoothing of the volume fraction $f_{l_1}$ is performed in each cell by utilising the tensor product with a quadratic B-Spline like suggested by Brackbill~et.~al.~\cite{BRACKBILL1992}. The smoothing can be repeated in multiple smoothing steps. Depending on the number of smoothing steps, more layers of neighbouring cells are included. One smoothing step was sufficient for the results presented in Sec.~\ref{Sec:Results}. For two-phase flows, the interface density 

\begin{equation}
   a_{\Gamma,l_1} = \frac{A_{\Gamma,l_1}}{V_{cell}} \approx || \mathbf{\nabla} f_{l_1} ||\text{,} \label{Eq:agamma}
\end{equation}
which indicates the amount of interface area $A_{\Gamma,l_1}$ in each cell, is approximated with the length of the interface's normal, which provides a formulation of the interface density inside the smoothed field. This approximation leads to a more smooth representation of the interfacial forces \cite{LAFAURIE1994} than utilising the highly discretisation dependent areas of the PLIC planes. The smoothed interface density $\tilde{a}_{\Gamma,l_1}$ is obtained from the application of Eq.~(\ref{Eq:agamma}) to the smoothed volume fraction field $\tilde{f}_{l_1}$.\\
From this baseline, the challenges arising for immiscible liquids are discussed in the upcoming Sec.~\ref{Sec:Problem}. New approaches to solve those problems are proposed in Sec.~\ref{Sec:3vof} and Sec.~\ref{Sec:CSS3ph}.
\section{Immiscible Liquids in VOF Simulations} \label{Sec:CLVOF}
\subsection{Problem Description} \label{Sec:Problem}
\begin{figure}
\centering
\begin{minipage}[t]{0.45\linewidth}
\centering
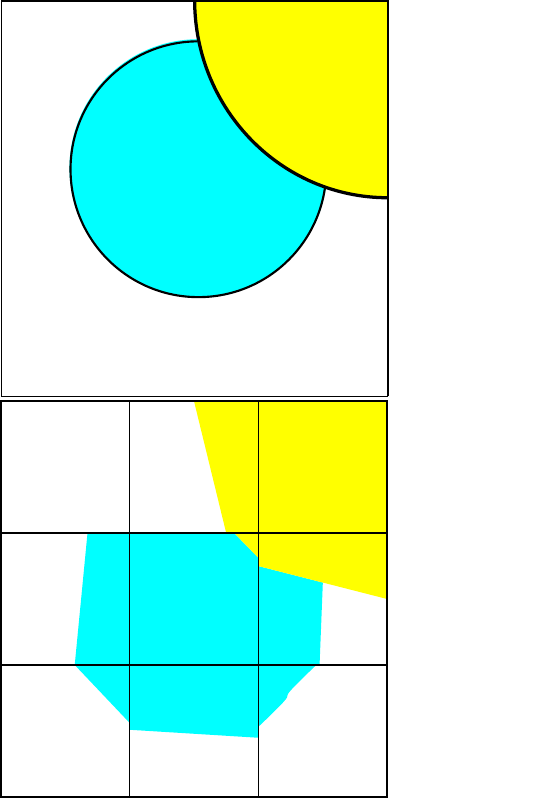
\caption{Exemplary distribution of two immiscible VOF variables and their desired interface reconstruction.}\label{Fig:ImmiscibleExample}
\end{minipage}
\hspace{0.04\linewidth}
\begin{minipage}[t]{0.45\linewidth}
\centering
  \begingroup%
  \setlength{\unitlength}{5.5cm}%
  \begin{picture}(1,1.47594952)%
    \put(0,0){\includegraphics[width=\unitlength,page=1]{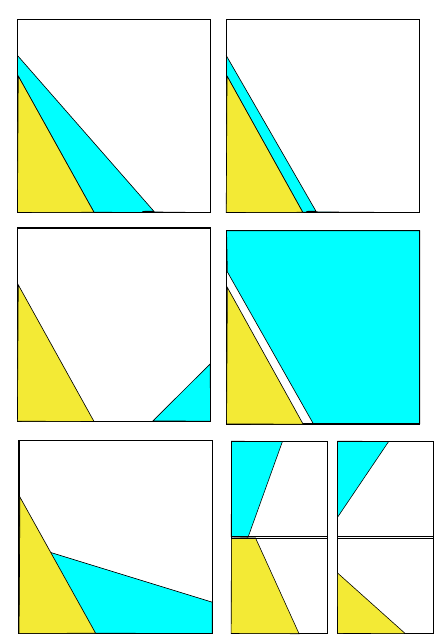}}%
    \put(0.35,1.33){(a)}%
    \put(0.83,1.33){(b)}%
    \put(0.35,0.83){(c)}%
    \put(0.83,0.83){(d)}%
    \put(0.35,0.36){(e)}%
    \put(0.637,0.36){(f)}%
    \put(0.873,0.36){(g)}
  \end{picture}%
\endgroup%
   \caption{Possible three component topologies of liquid~1 (yellow), liquid~2 (blue) and gas with and without a contact line.}
   \label{Fig:Threephasesituations}
\end{minipage}
\end{figure}
Starting with the VOF method with a PLIC interface reconstruction and the CSS model revised above for two-phase flow, alterations for the accurate modelling of immiscible droplet collisions are required. Methods for the interface reconstruction, advection and surface force computation of three-component situations are demanded, as the normal's computation according to Eq.~(\ref{Eq:GradientNormal}) as well as the interface density in Eq.~(\ref{Eq:agamma}) rely on the gradient of the volume fractions. The approximation of this gradient as well as the previously described smoothing require a sufficient stencil. In immiscible droplet collisions, a contact line occurs at the junction of the three liquid-liquid- and liquid-gas interfaces, analogous to three-phase flows where also a contact line is formed. In such situations, the gradient of the volume fraction approximated with a central difference does not yield the correct interface orientation as the volume fraction field is truncated for the liquid~2 (index $l_2$) which is spreading over liquid~1. This is depicted in Fig.~\ref{Fig:ImmiscibleExample}. It is clear, that a similar problem is present for three-phase flow \cite{PATHAK2016, PATEL2017}, thus the methods for the interface reconstruction proposed below will be applicable to both immiscible liquids' interaction as well as three-phase flow.\\
The correct geometrical reconstruction of the interface is only the basis to simulate the highly dynamic process of collisions. The reconstruction of the interface affects the advection as well as the surface force computation with the CSS model, because the smoothed interface orientation and topology information is employed inside the smoothing stencil there.\\
An other challenge in immiscible droplet collisions is, that thin films occur. ``Thin'' in a numerical sense means that the film thickness of the covering liquid or a portion of entrapped gas is below one cell's height. In a Cartesian grid, this situation always appears for contact angles other than $90^\circ$. Therefore, thin films of the encapsulating liquid (for contact angles below $90^\circ$) or the gas (for contact angles above $90^\circ$) are usually present in physical cases. A similar reconstruction problem arises, when the interfaces of both liquids reside in the same cell and it is unknown from the volume fractions alone, if one droplet already touches the other one and setting a contact angle is appropriate, or if the two interfaces are not yet wetting each other. \\ 
Proceeding from two-phase flow to two immiscible disperse liquids in a gaseous continuous phase, all above presented methods for two phases are applicable to each of the two immiscible liquids, as long as the two liquids stay adequately apart. As soon as the two liquids reside in the same cell and soon after that usually touch each other, some additions are required to account for the new liquid-liquid interface. Especially the contact line formed at the junction of the three interfaces of liquid~1-gas, liquid~2-gas, and liquid~1-liquid~2 needs additional modelling like indicated before. The subsequent Sec.~\ref{Sec:3vof} presents a topology capturing PLIC method as well as an adaption of the advection for three VOF variables with a sharp interface. Section~\ref{Sec:CSS3ph} shows an extension of the CSS model to three deforming interfaces for immiscible liquid interaction with thin covering films.
\subsection{Volume of Fluid Method for Three Components}
\label{Sec:3vof}
\begin{figure}
\centering
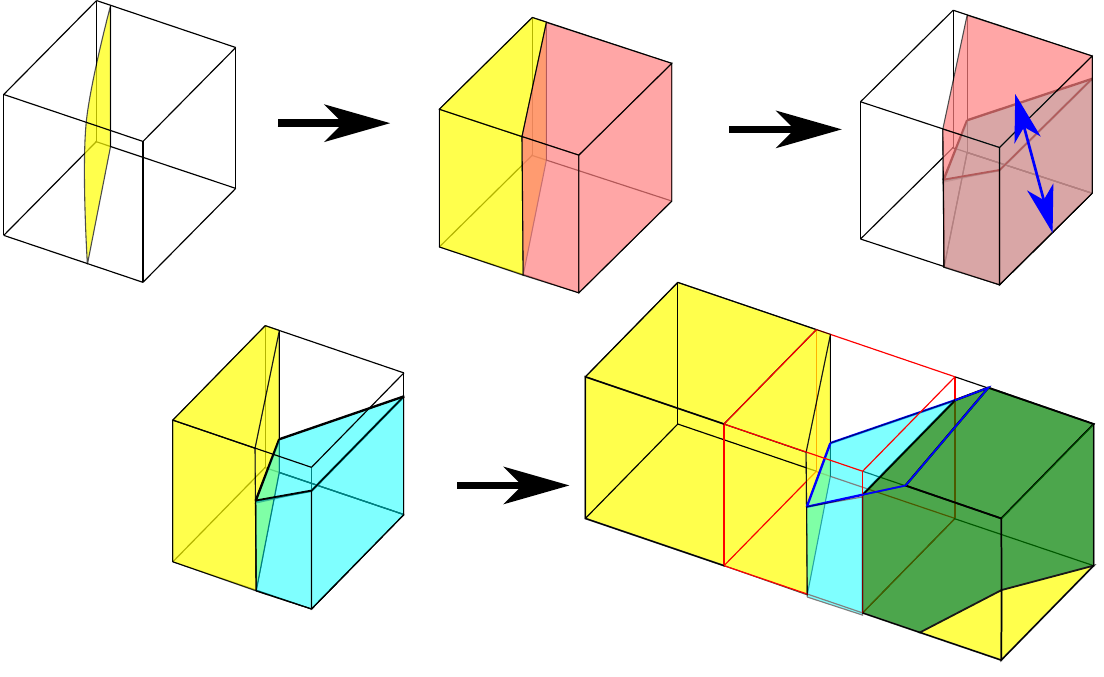
\caption{The position of the liquid~2's (blue) interface is found inside the remnant polyhedron (red) after cutting the cell with the liquid~1's (yellow) interface. The predicted volume fractions $f_p$ (green) result from an extension of the plane with orientation $\mathbf{n}_{l_2}$ and position $l_{l_2}$ to all neighbouring cells. The predicted volume fraction $f_p$ is calculated from the enclosed volume. During the extension, the distribution of liquid~1 inside the neighbouring cells must be considered during the volume computation. The cell's predicted volume fraction $f_{p,cell}$ on the other hand is the whole volume fraction enclosed in the neighbouring cell without subtracting the enclosed volume occupied by liquid~1.}
\label{Fig:gerror}
\end{figure}

An extension of the VOF method to immiscible liquids is performed largely analogous to three phases, e.g. \cite{PATHAK2016, PATEL2017, GOEHL2018, WASHINO2010}, except for the additional deformable interfaces. A new volume fraction $f_{l_2}$, cf. Eq.~(\ref{Eq:VOFVar}), of the second liquid is introduced. This second VOF variable is advected analogously to Eq.~(\ref{Eq:fadvect}) where the velocity $\mathbf{u}$ is equal to the cell average velocity for both liquids which facilitates the advection, but does not allow slip between the two liquids. The cell averages of density and dynamic viscosity 

\begin{equation}
\begin{split}
\rho_{cell} = f_{l_1} \rho_{l_1} + f_{l_2} \rho_{l_2} +\left(1-f_{l_1}-f_{l_2}\right) \rho_{gas} \\
\mu_{cell} = f_{l_1} \mu_{l_1} + f_{l_2} \mu_{l_2} +\left(1-f_{l_1}-f_{l_2}\right) \mu_{gas}
\end{split}
\label{Eq:PropertiesImmiscible}
\end{equation}
are still volume weighted averages, cf. Eq.~(\ref{Eq:Properties}).
\paragraph{Reconstruction} Reusing the idea by Focke and Bothe~\cite{FOCKE2011} of stabilising a film in combination with the idea for three-phase PLIC of Pathak and Raessi \cite{PATHAK2016} as well as Patel~et.~al.~\cite{PATEL2017}, a topology capturing modified PLIC algorithm for three components is proposed in the following. It should be noted, that for the reconstruction the only assumption is, that the volume fractions of two disperse components are known in each cell of the computational domain. This allows the application of this algorithm also to three-phase flow. \\
For all discussions in the following, it is assumed, that liquid~1 always has the higher surface tension at the interface towards the gaseous phase. This assumption complies with the findings from experiments, cf. \cite{CHEN2006,PLANCHETTE2009,PLANCHETTE2011,PLANCHETTE2012}, where the liquid with the higher surface tension is always covered by the one with the lower surface tension. The encapsulated, inner interface, the one of liquid~1, is reconstructed independently from liquid~2 and thus in analogy to the original PLIC for two phases discussed in Sec.~\ref{Sec:Basics}. The interface of liquid~2 is reconstructed afterwards, leading to a sequential PLIC algorithm. Therefore, the reconstruction of the liquid~2's interface is the focus of a new topology capturing PLIC method. \\
Different three-component situations are possible. Each topology requires different reconstruction methods. The upcoming explanations first discuss how a known configuration can be reconstructed. Later on it is addressed, how to choose the topology which fits best, desirably with low computational efforts. The following possible topologies in three-component cells and in the vicinity of the contact line are shown in Fig.~\ref{Fig:Threephasesituations}:
  \begin{itemize}
  \item[(a)] Fully wetting, different orientation of the interfaces.
  \item[(b)] Fully wetting, thin parallel layer of liquid~2.
  \item[(c)] No wetting, independent interfaces.
  \item[(d)] No wetting, with a very thin layer of gas and anti-parallel interfaces.
  \item[(e)] Situation with a contact line and arbitrary contact angle $\theta$.
  \item[(f)] Two neighbouring cells with a contact line at the cell border.
  \item[(g)] The cell contains only one liquid, both liquids occur in a $3\times3\times3$ stencil, but there is no contact line at the central cell's faces.
\end{itemize}
Like shown in multiple applications of the VOF method for three phases \cite{GOEHL2018, PATEL2017, PATHAK2016}, a contact angle representing the wetting behaviour is set for the case (e) and (f) with a contact line. This is neither appropriate for the fully wetting situation in (a) and (b) nor for the non-wetting topology in (c) and (d). Which topology is present is unknown from the volume fraction distribution at first. In the following explanation it is first assumed, that the topology is known for the reconstruction. The distinction of the correct topology is described later on. The previously discussed possible topology situations (a)-(g) can be reconstructed as follows:
\begin{itemize}
  \item[(a)] For the fully wetting case with different orientations of the interfaces, the interfaces are layered, but have different orientations of the normals. The liquid~2's interface has the normal
\begin{equation}
\mathbf{n}_{l_2} = \mathbf{\nabla} \left(f_{l_1}+f_{l_2}\right)
\end{equation}
which encloses the volume fraction $\left(f_{l_1}+f_{l_2}\right)$.
\item[(b)] For thin films covering the inner liquid~1, parallel layers of the two interfaces occur. The interface of liquid~2 with the normal
\begin{equation}
  \mathbf{n}_{l_2} = \mathbf{n}_{l_1}
\end{equation}
is again positioned such, that the volume corresponding to $(f_{l_1}+f_{l_2})$ is enclosed. This is similar to the ``onion skin" method by Youngs \cite{YOUNGS1982,BENSON2002}.
\item[(c)] If both interfaces reside in the same cell, but they do not touch, therefore do not wet each other, the interfaces are reconstructed independently. The liquid~2's interface normal
\begin{equation}
\mathbf{n}_{l_2} = \mathbf{\nabla} f_{l_2}
\end{equation}
as well as the positioning with the enclosed volume fraction $f_{l_2}$ is computed analogously to the original two-phase PLIC.
\item[(d)] When the two layers just do not touch, the gas film becomes very thin and the interfaces are anti-parallel to each other. Thus the liquid~2's interface normal is
\begin{equation}
\mathbf{n}_{l_2} = -\mathbf{n}_{l_1}
\end{equation}
and the position is set such, that it encloses the volume fraction $f_{l_2}$.
\item[(e)] The situation with an intersection of the interfaces inside the cell requires the setting of a contact angle $\theta$. As already mentioned above, there are different approaches in literature for three-phase reconstruction, which are applicable to immiscible liquids as they are also represented by two separated volume fraction distributions $f_{l_1}$ and $f_{l_2}$. An explicit approach to set a contact angle is shown in Patel~et~al.~\cite{PATEL2017} which is based on an idea of Washino~et~al.~\cite{WASHINO2010}. This idea is similar to the computation of the interface normals in the immersed boundary approach by Sussman \cite{SUSSMAN2001}. At first, the gradient of the liquid~2's volume fraction is projected onto the plane of the liquid~1's interface and normalised as

\begin{equation}
 \mathbf{t} = \frac{\mathbf{\nabla} f_{l_2} - (\mathbf{n}_{l_1}\cdot \mathbf{\nabla} f_{l_2}) \mathbf{n}_{l_1}}{|| \mathbf{\nabla} f_{l_2} - (\mathbf{n}_{l_1}\cdot \mathbf{\nabla} f) \mathbf{n}_{l_1}||}
\end{equation}
to extract components of the normal's orientation of the liquid~2's interface in the liquid~1's plane. Subsequently, the second orientation towards the interface of liquid~1 is set by imposing a known contact angle $\theta$
\begin{equation}
\mathbf{n}_{l_2} = \mathbf{n}_{l_1} \cos(\theta) + \mathbf{t} \sin(\theta)
\end{equation}
to orient the interface's normal. This contact angle is prescribed in simulations. For many liquid combinations the static contact angle is known from measurements. For a moving contact line, $\theta$ is a dynamic contact angle. For liquid-solid interaction various correlations exist in literature, examples can be found in \cite{KISTLER1993, JIANG1979, COX1986, BLAKE2002, TANNER1979}, but for immiscible liquids in air this is still an open research question. The positioning of the interface of the liquid~2 is performed inside the remnant polyhedron, after the cuboid cell is cut with the interface of liquid~1. This is illustrated in Fig.~\ref{Fig:gerror} (upper row). An efficient sequential positioning of two interfaces is described by Kromer~et.~al.~\cite{KROMER2021}. After the positioning, it is checked, if the interfaces really result in a wetting situation, otherwise the cell is marked as ``non-wetted" and a further improvement of the orientation can be performed which is discussed later on.
  \item[(f)] The contact angle is also set for two neighbouring cells with a contact line at the cell border. Cells with a potential contact line are identified as cells, where liquid~1 is inside the liquid~2's interface reconstruction stencil and additionally liquid~1 in a neighbouring cell wets the common cell face. The setting of the contact angle is performed analogously to case (e) with a contact line inside the cell. The positioning is analogous to the two-phase case, because there is only one volume fraction inside the cell. If the interfaces really result in a wetting situation, is checked after the calculation of the PLIC interface with the initial assumption of a contact line. If the comparison of the true to the assumed topology does not hold, a ``non-wetted" situation is found, the cell is marked as such and a correction can be performed in a later step with a virtual VOF field. This is explained below in this section.
\item[(g)] If the cell contains only one liquid, but both liquids occur in a $3\times3\times3$ stencil, the field for the normal's calculation is truncated and again a special case arises for the reconstruction. These cells are also marked as ``non-wetted" and a correction of the PLIC positioning will be performed in a later step with a virtual VOF field described below.
\end{itemize}
A method is needed to identify which reconstruction is the real topology. For this, the idea by Pathak and Raessi \cite{PATHAK2016} of calculating an error for a given orientation is reused here. Pathak and Raessi propose the calculation of the error according to

\begin{equation}
    g_{i,j,k}(\mathbf{n}_{l_2}) = \sum\limits_{i^*=i-1}^{i+1}\sum\limits_{j^*=j-1}^{j+1}\sum\limits_{k^*=k-1}^{k+1} (f_{l_2,i^*,j^*,k^*}-f_{p,i^*,j^*,k^*}(\mathbf{n}_{l_2}))^2 \label{Eq:gerr}
  \end{equation}
as a measure for the quality of the reconstruction with an arbitrary normal $\mathbf{n}_{l_2}$. The predicted volume fraction $f_p$ and true volume fraction $f_{l_2}$ are compared inside the $3 \times 3 \times 3$ stencil of all neighbouring cells. The predicted volume fraction $f_p$ is found in each neighbouring cell $(i^*,j^*,k^*)$ by extending the PLIC plane in the present cell $(i,j,k)$ to its neighbours in a $3 \times 3 \times 3$ stencil. The extension to the neighbouring cells is visualised exemplary in Fig.~\ref{Fig:gerror}.
This idea of an extension to all neighbours is adopted here to decide which of the orientations (a)-(e) above is the best representation of the true topology instead of performing a costly minimisation throughout all orientations of the unit sphere's interface like Pathak and Raessi suggest. The following procedure is applied: 
\begin{enumerate}
\item As a first criterion for the preferred orientation, the number of completely filled neighbouring cells is counted where both the predicted ($f_p=1$) and true volume fraction ($f_{l_2}=1$) are unity. The topology with the highest count is chosen, if criterion 3 is not violated. If the counts are equal, criterion 2 is employed for the final decision.
\item If criterion 1 identifies multiple orientations with equal counts, Eq.~(\ref{Eq:gerr}) is employed as a second criterion for the decision, which orientation reflects the real topology. Additionally, criterion 3 should again not be violated.
\item In addition to criterion 1 and 2, a criterion is necessary to enforce a contact angle wherever a contact line is formed. The orientation is not chosen, if one of the following criteria identify a topology mismatch between the assumed and resulting topology. A topology mismatch is identified,
\begin{itemize}
\item if a contact line is present for an orientation which assumes no contact line. This is tested by comparing the enclosed volume fraction $f_{p,cell}$ (cf.~Fig.~\ref{Fig:gerror}) below the PLIC plane without the assumption of a contact line to the true volume fraction $f_{l_2}$. A mismatch is found, if $f_{p,cell}$ is neither equal to $f_{l_2}$ for the non-wetted case (c) nor $f_{l_1}+f_{l_2}$ for the fully wetted case (a).
\item if liquid~2 does not enclose liquid~1 for an assumed liquid film stabilisation case (a). The comparison is done analogous to the previous case. Additionally, for the liquid film stabilisation case the extension to the neighbouring cells should result in predicted volume fractions above zero, if there is liquid~2 inside that cell.
\item if the cells, where the interface extends to, has no liquid~2 inside for an assumed gas- or liquid film stabilisation topology.
\end{itemize}
\end{enumerate}
This additional checks in step 3 are computationally inexpensive, as all values needed are already computed. If the contact angle configuration yields the best orientation, but has no contact line, the cell is marked as ``non-wetted" for additional refinement of the orientation with the aid of a virtual VOF field later on. The parallel alignment cases (b) and (d) are only tested, if (a) or (c) are the best orientation regarding to Eq.~(\ref{Eq:gerr}), but result in the formation of a contact line. Up to here, usually three, at most five evaluations, if the anti-parallel and parallel cases are tested, of Eq.~(\ref{Eq:gerr}) are necessary. One additional evaluation will be necessary for the three-component cells which are marked as ``non-wetted".\\
As indicated before, a virtual VOF field $f_{l_2,virt}$ is needed to orientate and position the PLIC planes in cells which are marked as ``non-wetted". For the virtual VOF field the interfaces of liquid~2 in cells with a contact line are extended geometrically into liquid~1 to avoid truncated volume fraction values in central stencils used for the normals' computation in ``non-wetted" cells. For the reconstruction of ``non-wetted" cells one additional layer of liquid~2 volume fractions is needed. In Sec.~\ref{Sec:CSS3ph} it is shown that the additional values required depend on the number of smoothing steps employed for the surface force computation. The optimal results for the resolution of the simulations shown in Sec.~\ref{Sec:Results} were obtained with one smoothing step, thus two layers of virtual VOF values, although the implementation was tested for up to four layers. As it is impractical regarding the performance to intersect every cell with all existing interfaces which form the contact line inside the whole computational domain, especially if a large domain and a parallelisation with distributed memory is considered, the search stencil in which the contact line is looked for is adapted to the contact angle. Figure~\ref{Fig:SearchStencil} shows the justification in simplified two dimensional examples for different contact angles. For fully wetting liquids or very low contact angles this search stencil is limited, as it otherwise reaches throughout the whole computational domain. In this study a maximum of four cells, thus a maximum central stencil of $9\times9\times9$, was sufficient for the results in Sec.~\ref{Sec:Results}.
\begin{figure}
\centering
\null\hfill
\begin{minipage}[t]{0.65\linewidth}
\includegraphics[width=\linewidth,page=1]{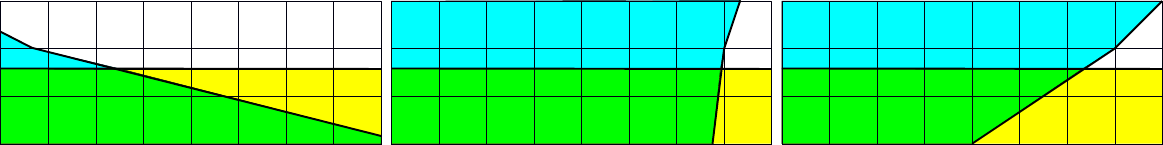}
\caption{If the contact angle is very low or close to $180^\circ$, the search stencil for the virtual volume fraction (green) is the largest, while close to $90^\circ$ it only has to be extended in one direction. Here, the row of virtual values in three component cells and one additional row of virtual VOF values inside liquid~1 are depicted, which is sufficient for a maximum of one smoothing step. The depth of the extension has to be adapted to the number of smoothing steps used in the CSS model.} 
\label{Fig:SearchStencil}
\end{minipage}
\hfill
\begin{minipage}[t]{0.3\linewidth}
\centering
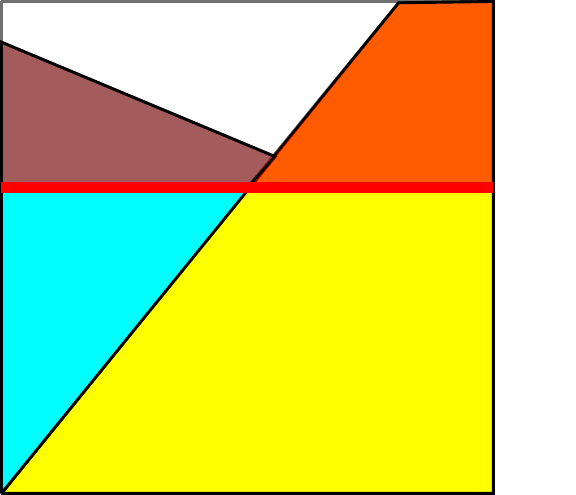
\caption{For the advection the two polyhedrons are computed with the PLIC interfaces and subsequently cut with a plane parallel to the cell's face the flux is computed for, depicted in a two-dimensional, thus simplified, form for the $y$-direction with velocity component $v$ and time step $\Delta t$.}
\label{Fig:Advection3ph}
\end{minipage}
\hfill\null
\end{figure}
For the two-component cells (g) which are initially marked as ``non-wetted", the virtual VOF field is now used to compute the normal with Eq.~(\ref{Eq:GradientNormal}). The position of the plane is again analogous to two-phase PLIC. Additionally, for the three-component cells and two-component cells where the contact angle orientation was tested, but later on marked as ``non-wetted", again criteria 1 to 3 are employed to test, if this orientation from the gradient of the virtual VOF field will represent the true topology even better. \\
In the very rare worst case that all orientations are tested and additionally marked as ``non-wetted" this explicit algorithm leads to at most six evaluations and in the most common cases only three evaluations of Eq.~(\ref{Eq:gerr}). This is in any case less than the ten initial bracketing evaluations and an additional minimisation required in the algorithm by Pathak and Raessi \cite{PATHAK2016}. The algorithm is at the same time able to capture the topology of immiscible droplet collisions as the results in Sec.~\ref{Sec:Results} show. It should be noted, that also without this separate evaluation of ``non-wetted" cells after the computation of the virtual VOF field, the interface reconstruction is satisfactory using only the orientations above. As the CSS model for immiscible liquid interaction presented below needs the virtual VOF field, this was also included in a refinement of the interface reconstruction in this study. After the reconstruction of the interfaces the advection is performed.
\paragraph{Advection of Three Immiscible VOF Variables}
When the topology capturing reconstruction with a contact line and thin films is completed, the split advection is performed. The two reconstructed components' interfaces form two convex polyhedrons. Those are cut with a plane parallel to the adjacent face of the current advection direction. For this purpose the polyhedrons of each phase are computed and subsequently the intersections of them with the mentioned plane. This requires noticeable computational effort for the retrieval of the connectivity of each polyhedron after each cut for the volume computation. The volume fluxes are exemplary illustrated in two dimensions for one direction in Fig.~\ref{Fig:Advection3ph}.
\subsection{Surface Force Modelling for Three Immiscible Components} \label{Sec:CSS3ph}
The calculation of the surface forces in the vicinity of a contact line with the CSS model requires further modelling of the additional deformable interfaces for immiscible liquid interaction. For two liquids inside the continuous gas phase three different interfaces occur: 
\begin{itemize}
\item[(1)] The liquid~1 ($l_1$) - gas ($g$) interface with the interfacial tension $\sigma_{l_1-g}$
\item[(2)] The liquid~2 ($l_2$)- gas interface with the interfacial tension $\sigma_{l_2-g}$
\item[(3)] The liquid~1 - liquid~2 interface with the interfacial tension $\sigma_{l_1-l_2}$
\end{itemize}
An extension of the CSS model to immiscible liquids is facilitated by introducing partial surface tensions $\gamma_i$ for each component i. Following the idea of Smith~et~al.~\cite{SMITH2002} and Joubert~et.~al.~\cite{JOUBERT2020}, the interfacial tensions between component $i$ and $j$, 

\begin{equation}
\sigma_{ij} = \gamma_i + \gamma_j,
\label{Eq:partialsigma}
\end{equation}
are described as sums of the partial interface tensions $\gamma_i$, with $i$ and $j$ being the adjacent components of an interface. The tensor

\begin{equation}
\mathbf{T}_i = -\mathbf{\nabla} \cdot \left(\tilde{a}_{\Gamma,i} \left(\mathbf{I}-\tilde{\mathbf{n}}_{\Gamma,i} \otimes \tilde{\mathbf{n}}_{\Gamma,i}\right)\right)
\label{Eq:T}
\end{equation}
is used for the superposition of the interfaces' contributions shown below in Eq.~(\ref{Eq:Superposition}). The smoothed volume fraction fields $\tilde{f}_i$ are obtained analogously to the two-phase case with applying the quadratic B-Spline smoothing operator to the volume fraction field. It should be noted, that up to here, the volume fraction of the gas

\begin{equation}
f_g = 1 - f_{l_1} - f_{l_2}
\end{equation}
was not computed yet, but is required here. For liquid~2, the virtual volume fraction field $f_{l_2,virt}$ is utilised to provide values inside the truncated smoothing stencil close to three-component cells. The overall acceleration acting in the vicinity of the interface $\Gamma$

\begin{equation}
\left(\frac{\partial \mathbf{u}}{\partial t}\right)_\Gamma =\frac{1}{\rho} \sum\limits_{i=1}^{3}\gamma_i \mathbf{T}_i \text{.}
\label{Eq:Superposition}
\end{equation}
results from the superposition and a scaling with the density $\rho$, which is chosen dependant on the interfaces present which is discussed below. The orientations $\mathbf{\tilde{n}}_{\Gamma,i}$ and the interface densities $\tilde{a}_{\Gamma,i}$ are computed with Eq.~(\ref{Eq:GradientNormal}) and Eq.~(\ref{Eq:agamma}) for the respective volume fraction. For liquid~2 the virtual volume fraction field described in Sec.~\ref{Sec:3vof} already for the reconstruction is employed. The information of the interfaces present is necessary for the choice of the density according to

\begin{equation}
\rho =
\begin{cases}
0.5 \cdot (\rho_{l_1} + \rho_g), \text{in 3 component cells in the smoothed field}\\%
0.5 \cdot (\rho_i+\rho_j), \text{in 2 component cells with smoothed $i-j$-interface}%
\end{cases}
\label{Eq:RhoChoice}
\end{equation}
which scales the forces for the calculation of the acceleration due to the interfacial tensions, cf. Eq.~(\ref{Eq:Superposition}). As the topology information is not available inside the smoothed field, the information obtained from the reconstruction of the non-smoothed fields is used to identify cells with a thin film. For all other cells the information of the smoothed volume fractions is used to identify the interfaces present. The surface force modelling requires again a film stabilisation of thin liquid films. In cells with a thin liquid~2 film covering liquid~1 and cells with a contact line which are surrounded by such liquid film cells, the accelerations are calculated according to 

\begin{equation}
\left(\frac{\partial \mathbf{u}}{\partial t}\right)_{\Gamma,\mathrm{film}} =\frac{2}{\rho_g+\rho_{l_1}}  (\gamma_{l_1} + 2\gamma_{l_2} + \gamma_g ) \mathbf{T}_{l_1}
\label{Eq:accelfilm}
\end{equation}
to obtain this stabilisation. This modification exploits that the interfaces are (approximately or exactly) parallel to each other and the film is very thin. The covering liquid reduces the interfacial tension compared to the liquid~1-gas interface. However, the thin film does not act like a liquid-liquid interface yet. Therefore, a more suitable approximation of the surface force at a thin film is a scaling with the inner liquid's density jump towards the surrounding gas. This choice of the density strongly affects the collision outcome of two immiscible liquid droplets. The choice of the density in three-component cells according to Eq.~(\ref{Eq:RhoChoice}) and at a thin film according to the scaling density in Eq.~(\ref{Eq:accelfilm}) is motivated by a comparison to experimental data presented in Sec.~\ref{Sec:Results}. The comparison to experimental data on collisions of fully wetting liquids has shown that the modelling of the surface forces is very delicate. The number of smoothing steps required may be related to the resolution, as the results in Sec.~\ref{Sec:Results} indicate, but this finding will need further assessment beyond this work. Also, the choice of the scaling density, especially at the thin films, strongly affects the collision's outcome. Simulations of the partial wetting behaviour qualitatively yield the expected behaviour (compare Sec.~\ref{Sec:InterfacialTensionTests}) with this choice of the scaling density. No stabilisation of the thin gas film was considered here, but below a fraction of a cell height a thin gas film was handled the same like the liquids are in touch for the CSS model, as the effect of a gas film on the droplet collision outcome of immiscible liquids was insignificant in determining the coalescence and separation regimes. However, a stabilisation of the gas film may become relevant, if not the dominating influence besides modelling the pressure, if bouncing at very low velocities should be simulated. Unfortunately, sufficient validation data is not available in this regime. Summarised, the here presented methods are well suited for fully wetting immiscible liquids' interaction, as the following Sec.~\ref{Sec:Results} will prove. Possible solutions for partial wetting as well as collisions at low velocities were discussed here, but cannot be validated yet due to insufficient data in literature. In the following section, a validation for the collision of droplets of fully wetting liquids in different regimes and academic test cases show the potential of the here described methods to enable the simulation of three-dimensional immiscible droplets' interaction.

\section{Validation and Results} \label{Sec:Results}
Different test cases with increased complexity are discussed in the following. Test cases for the reconstruction of thin films of the covering liquid or the gas as well as situations with a contact line with different contact angles are discussed in Sec.~\ref{Sec:ReconstructionTests}. A convergence study is shown for the advection in Sec.~\ref{Sec:AdvectionTests}. Afterwards, test cases for the suggested modification of the CSS model for immiscible liquids are presented in Sec.~\ref{Sec:InterfacialTensionTests}. A comparison of the simulations with the new algorithms in FS3D to experimental data shows excellent agreement.
\subsection{Reconstruction} \label{Sec:ReconstructionTests}
\begin{figure}
\null\hfill
\begin{minipage}[b]{0.3\linewidth}
\centering
\includegraphics[width=0.5\textwidth, trim = 9cm 1cm 9cm 1cm, clip]{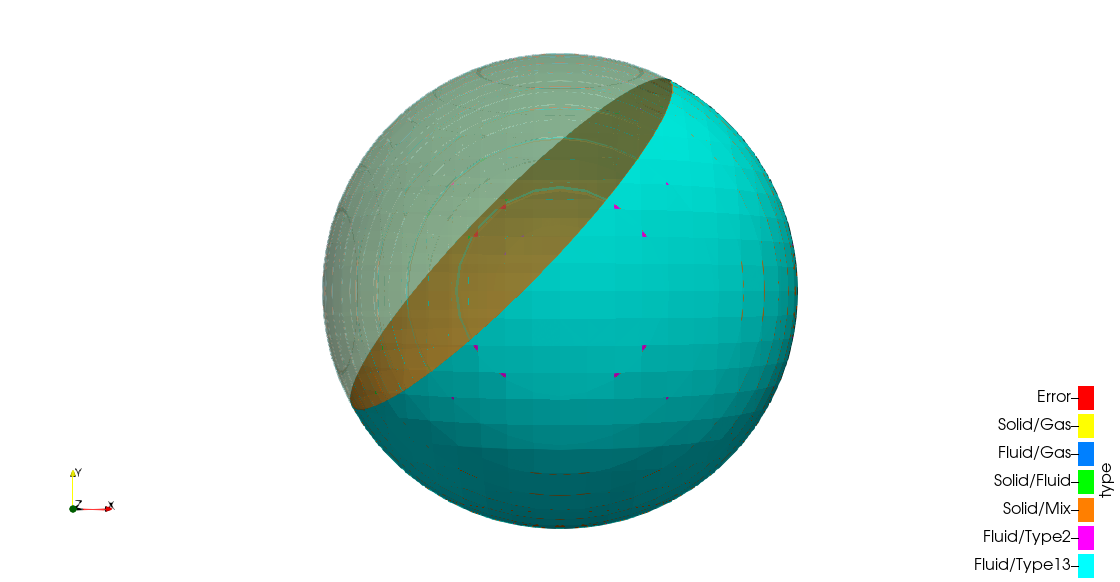}
\caption{Case 1: A spherical droplet encapsulated by a thin film of an other liquid. A part of the compound droplet is shown transparent. A contact angle of $\theta=60^\circ$ is used for the evaluation of the normal with a contact line, but this orientation is never set. The diameters present are $d_{l_1} = 6.62~\mathrm{mm}$ and $d_{l_2}=6.7~\mathrm{mm}$ in a $(10~\mathrm{mm})^3$ domain resolved with $32$ cells per direction.} \label{Fig:Case1ThinFilm}
\end{minipage}
\hfill
\begin{minipage}[b]{0.24\linewidth}
\centering
\includegraphics[width=0.8\textwidth, trim = 0cm 1cm 5cm 1cm, clip]{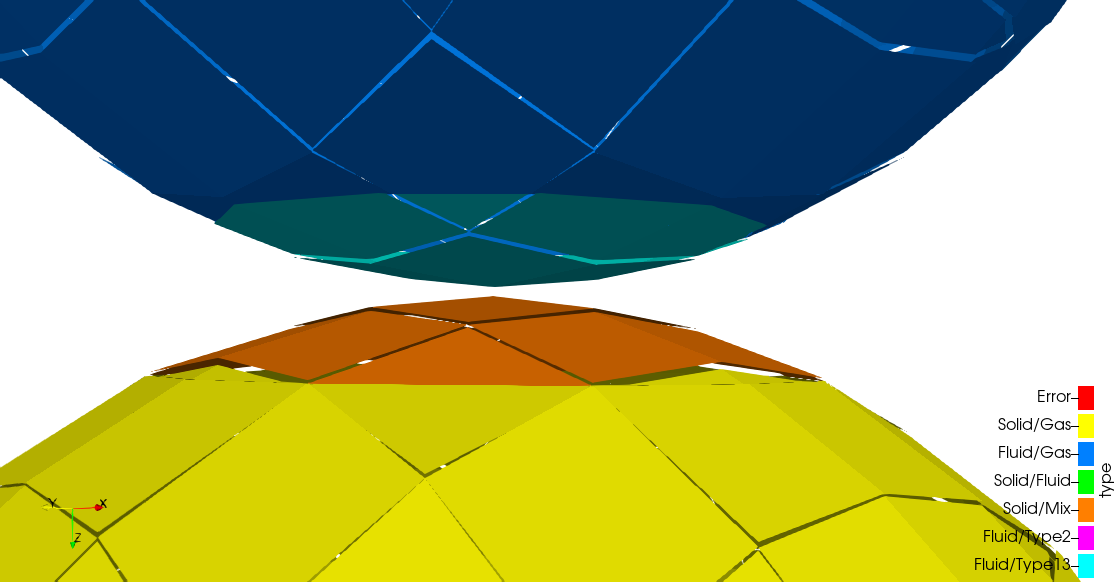}
\caption{Case 2: The two liquids' interfaces reside in the same cell, but do not touch. The diameters of the two droplets are equal ($3$ mm), the contact angle utilised in the algorithm is $60^\circ$, the domain is $(10 \mathrm{mm})^3$ resolved with $32$ cells per direction. \vspace{\baselineskip}}\label{Fig:Case2Gasfilmstabil}
\end{minipage}
\hfill
\begin{minipage}[b]{0.36\linewidth}
\centering
\includegraphics[width=\linewidth, trim = 6cm 0cm 6cm 1cm, clip]{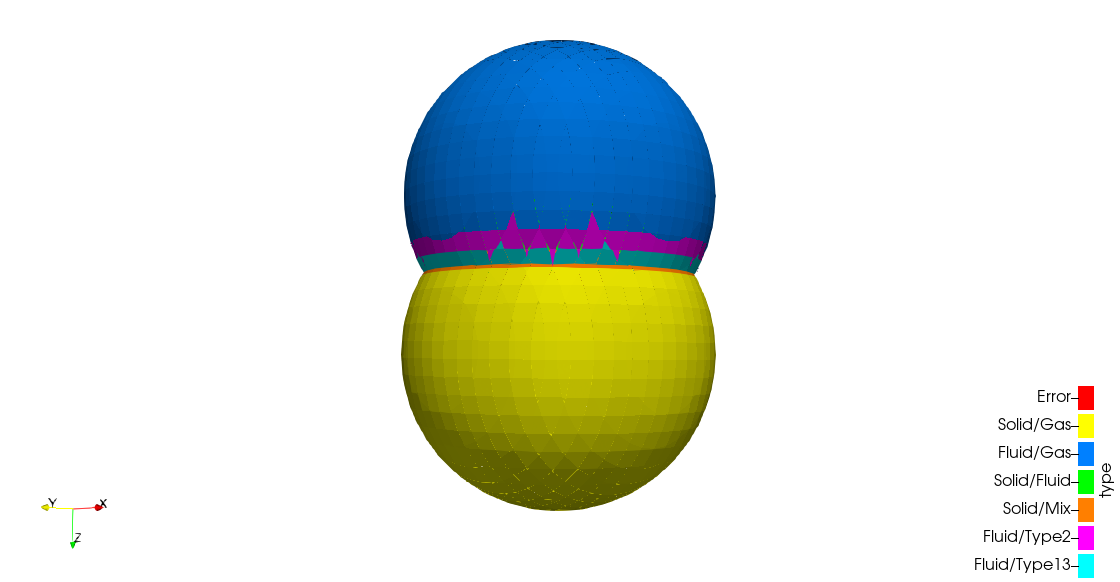}
\caption{Case 4: $\theta = 60^\circ$. The diameters of the droplet and the spherical cap are equal ($3$ mm), the contact angle utilised in the algorithm is $60^\circ$, the domain is $(10 \mathrm{mm})^3$ resolved with $64$ cells per direction.}
\label{Fig:Case4Theta60deg}
\end{minipage}
\hfill\null
\caption*{Figures \ref{Fig:Case1ThinFilm}-\ref{Fig:Case4Theta60deg}: Test cases for the reconstruction of the interface: The interfaces of both liquids (yellow for the interface of liquid~1, blue for liquid~2) which are in the same cell or with both liquids inside the reconstruction stencil, where the modified topology capturing film stabilisation PLIC is used, are marked with other colors (orange for the interface of liquid~1, light blue for the interface of liquid~2, magenta for the cells where both liquids are inside the reconstruction stencil). All pictures from simulations show the PLIC planes visualised with the aid of a Paraview Plugin developed by VISUS at the University of Stuttgart (see: https://github.com/UniStuttgart-VISUS/tpf).}
\end{figure}

For all the reconstruction test cases, the full algorithm described in Sec.~\ref{Sec:CLVOF} is applied. It is shown, that for all of the following situations the geometry is reconstructed correctly with the topology capturing PLIC algorithm. Geometries which are typical for binary droplet collisions of immiscible liquids are tested: A thin liquid and gas film as well as situations with different contact angles.
\paragraph{Case 1: Thin film of an immiscible liquid on a spherical droplet}
Thin films usually occur in the collisions of two immiscible liquid droplets, as the contact angles are low or in the fully wetting case close to zero. Therefore, it is essential, that a thin film with a height below the edge length of one cell is reconstructed correctly by the PLIC algorithm described above. A sphere of liquid~1 is initialised which is encapsulated by liquid~2 with a film height well below a cell's edge length. Figure~\ref{Fig:Case1ThinFilm} shows that the encapsulation of the inner droplet of liquid~1 with a liquid~2 thin film is reconstructed correctly in this case.
\paragraph{Case 2: Thin gas film between two liquids}
When two droplets approach each other, a thin gas film is formed between those two droplets. Setting the contact angle in all cells with liquid~1, liquid~2 and gas present cannot account for this situation. The new topology capturing PLIC presented above which enables a film stabilisation reconstructs this situation correctly. The reconstruction of the two droplets' interfaces residing in the same cell but not yet in contact is shown in Fig.~\ref{Fig:Case2Gasfilmstabil}.

\paragraph{Cases~3-6: Reconstruction with a contact line}
The reconstruction with a contact line is validated with a sphere of liquid~1 where a spherical cap of liquid~2 resides on top. The distance 
\begin{equation}
h = sin\left(0.5~\theta\right)~d
\end{equation}
of the centre points of the two initialised spheres is set such, that the desired contact angle $\theta$ is formed by the spheres' intersection. The diameters $d$ are equal for the liquid~1 sphere and the spherical cap of liquid~2. The reconstruction algorithm described above is validated for different contact angles: The here discussed test cases are $\theta = 40^\circ$ (Case~3), $60^\circ$ (Case~4), $90^\circ$ (Case~5) and $150^\circ$ (Case~6). Figure~\ref{Fig:PhiPlot} shows the reconstructed circumferential orientation $\phi_{sim}$ compared to the orientation $\phi_{ref}$ of the line between cell's centre which contains the interface and the centre point of the sphere for the previously mentioned test Cases 3-6 with different contact angles at the same resolution.
\begin{figure}
\null\hfill
\begin{minipage}[t]{0.54\linewidth}
\centering
\begin{tikzpicture}[scale=1]
\begin{axis}[
				width=6cm,
				xmin=-3.16,
            	ymin=-3.16,ymax=3.4,
            	xtick={-3.142,-1.571,0,1.571,3.142},
            	xticklabels={$-\pi$,$-\pi/2$,$0$,$\pi/2$,$\pi$},
            	ytick={-3.142,-1.571,0,1.571,3.142},
            	yticklabels={$-\pi$,$-\pi/2$,$0$,$\pi/2$,$\pi$},           
            	grid = major,
            	scaled ticks=false,
            	axis lines = left,
            	axis equal = true,
            	xlabel={$\phi_{ref}$},
            	ylabel={$\phi_{sim}$},
            	scaled ticks=false,
		        line width=0.75pt,
				legend entries={Case 3: $\theta=40^\circ$, Case 4: $\theta=60^\circ$, Case 5: $\theta=90^\circ$, Case 6: $\theta=150^\circ$,reference},
				legend style={
					at={(1.2,1.15)},
					anchor=north,
					legend columns=1,
					cells={anchor=west},
					rounded corners=2pt,}
					]
					\addplot[blue,mark=x,only marks,mark size=1.5pt] table {./Txt-files/phi_theta40deg.txt};
\addplot[red,mark=x,only marks,mark size=1.2pt] table {./Txt-files/phi_theta60deg.txt};
\addplot[green,mark=x,only marks,mark size=1pt] table {./Txt-files/phi_theta90deg.txt};	
\addplot[cyan,mark=x,only marks,mark size=0.52pt] table {./Txt-files/phi_theta150deg.txt};				
				\addplot[black, mark=x, mark size = 0.1pt, line width =2pt] table {./Txt-files/Test.txt};
\end{axis}
\end{tikzpicture}
\caption{The circumferential angle $\phi_{sim}$ in all cells at the contact line and close to it (liquid~1 inside the central difference stencil) are compared to the analytical reference orientation $\phi_{ref}$ in the cell centre of the orientation of the PLIC planes at the contact line at different contact angles $\theta$. As the reference solution is always computed with the cell's centre, but in the simulation the interfaces are not always placed inside the cell centre, a small deviation from the reference solution is expected for a finite grid spacing.}
\label{Fig:PhiPlot}
\end{minipage}
\hfill
\begin{minipage}[t]{0.42\linewidth}
	\centering
  \begin{tikzpicture}[scale =1.0]
    		\begin{loglogaxis}[
    		    width = 6cm,
    	      	axis lines=left,
            	xlabel={number of cells per direction $N$},
            	ylabel={${adv}_{\mathrm{error}}$},
            	xmin=24,xmax=128,
            	ymin=0.001,ymax=0.1,
		    line width=1pt,
        	    grid=major,
            	scaled ticks=false,
	            minor tick num=1,
	            xtick={24,32,48,64,96,128},
	            xticklabels={24,32,48,64,96,128},
				legend entries={error liquid 1 , error liquid 2},
				legend style={
					at={(0.9,1.1)},
					anchor=north,
					legend columns=1,
					cells={anchor=west},
					rounded corners=2pt,}
				]
				\addplot [red,mark=o] table [x=N, y=uniterrorf3] {./Txt-files/Advection_Convergence_DropletSpericalcap.txt};
				\addplot [black,mark=x] table [x=N, y=uniterrorf] {./Txt-files/Advection_Convergence_DropletSpericalcap.txt};
		 	\end{loglogaxis}
     	\end{tikzpicture}
\caption{Convergence study for the advection: The average $L1$-error $adv_{\mathrm{error}}$ in the volume fractions $f_{l_1}$ and $f_{l_2}$ is shown after one revolution of Case~4. The number of cells $N$ is the amount of grid cells per direction of the Cartesian domain. An experimental order of convergence above one is reached.}
\label{Fig:ConvergenceAdvection}
\end{minipage}
\hfill\null
\end{figure}

\subsection{Advection} \label{Sec:AdvectionTests}
For the validation of the advection, exemplary the reconstruction Case~4 from the previous Sec.~\ref{Sec:ReconstructionTests} with a contact angle of $60^\circ$ is revolved once around an axis orthogonal to the line connecting the spheres' centre points. After one revolution the errors in the volume fractions are computed in each cell and averaged. A convergence study of this average error

\begin{equation}
adv_{\mathrm{error}} =\frac{\sum\limits_{cell=1}^{Ncells}(|| f_{init}(cell)-f(cell))||}{\sum\limits_{cell=1}^{Ncells} f_{init}(cell)}
\end{equation} is shown in Fig. \ref{Fig:ConvergenceAdvection}. $N_{cells}$ is the total number of cells in the computational domain, $f_{init}$ is the initialised volume fraction of the respective liquid and $f$ the volume fraction after one revolution. An experimental order of convergence slightly above 1 is reached. It should be noted, that the error for the reconstruction of the interface of liquid~2 is always higher than the one for liquid~1. This is expected, as the reconstruction is performed sequentially. Therefore, the error of the reconstruction of liquid~1 contributes to the reconstruction of the liquid~2's interface in three-component cells which can be seen also in the presented advection convergence study.
\subsection{Interfacial Forces}\label{Sec:InterfacialTensionTests}
The validation of the implementation of the CSS model's modifications for two immiscible liquids forming a contact line is shown here. For the validation of the wetting behaviour, two academic test cases which are initialised identically, are shown. A comparison to experimental results of near head-on coalescence and crossing separation as well as off-centre stretching separation is presented subsequently. 
\begin{table}[bt!]
\centering
\begin{tabular}{c c c}
\hline
Distance of droplet's centre points & $h$ & $50~\mathrm{\mu m}$ \\
Diameter liquid~2 spherical cap & $d_{l_2}$ & $200~\mathrm{\mu m}$ \\
Diameter liquid~1 sphere & $d_{l_1}$ &   $150~\mathrm{\mu m}$ \\
Density of the liquids & $\rho_{l_1}=\rho_{l_2} $ & $800~\mathrm{kg/m^3}$ \\
Dynamic viscosity of the liquids & $\mu_{l_1}=\mu_{l_2} $ & $5.0\cdot 10^{-3}~\mathrm{kg/(m~s)}$\\
Density of the air & $\rho_{\mathrm{air}}$       & $1.19~\mathrm{kg/m^3}$ \\
Dynamic viscosity of the air & $\mu_{\mathrm{air}}$ & $18.24 \cdot 10^{-6}~\mathrm{kg/(m~s)}$ \\
\hline
\end{tabular}
\caption{Geometry and liquids' properties for both fully and partial wetting test cases. The surface tensions are adjusted for the desired wetting behaviour.}  \label{Tab:GeometryProperties1}
\end{table}

\begin{figure}[tb]
\newcommand\filepath{.}
\null\hfill
\begin{subfigure}[t]{0.19\linewidth}
\centering
\includegraphics[scale=0.4, trim = 25cm 19cm 25cm 0cm,clip]{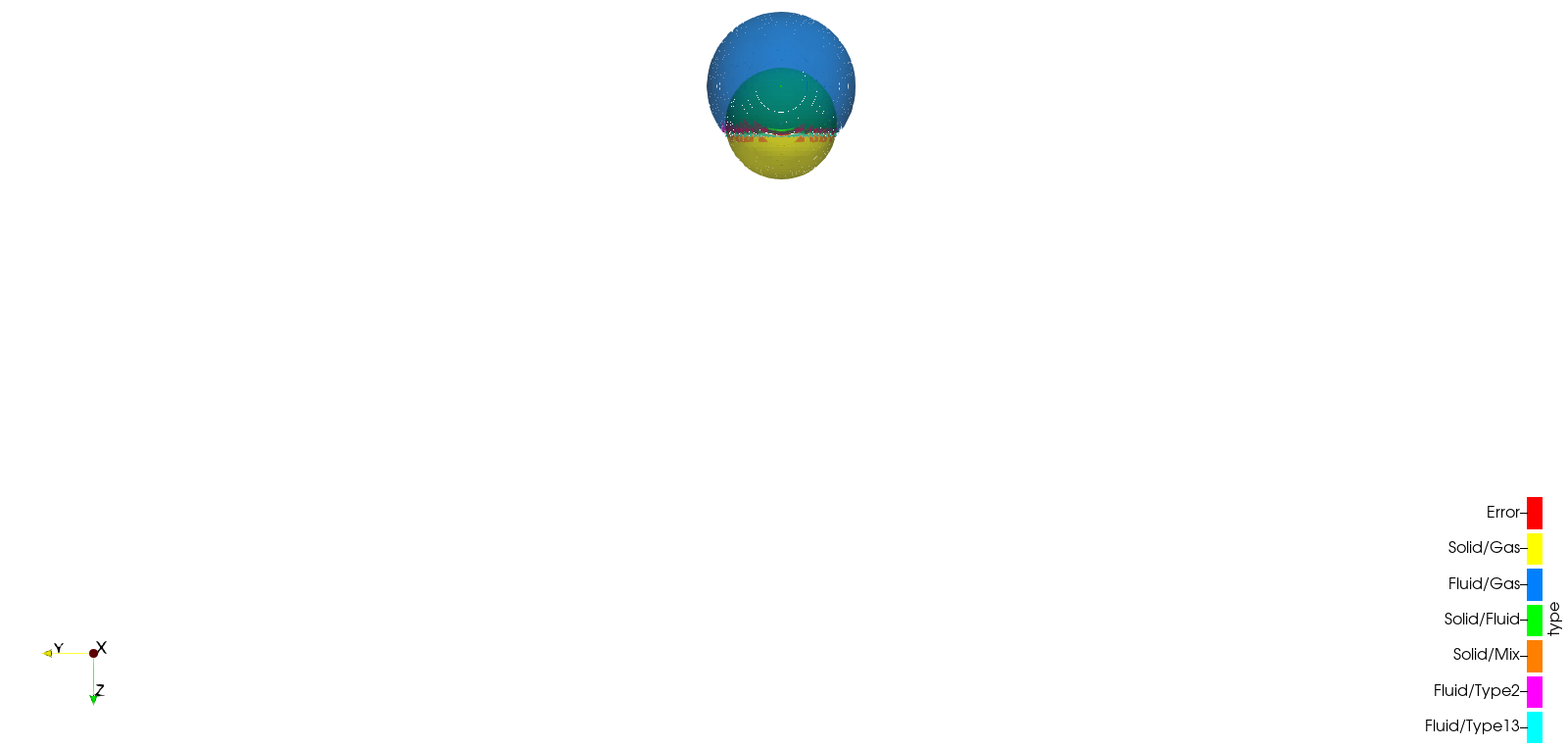}
\caption{$t=0~\mathrm{\mu s}$}
\end{subfigure}
\hfill
\begin{subfigure}[t]{0.19\linewidth}
\centering
\includegraphics[scale=0.4, trim = 25cm 19cm 25cm 0cm,clip]{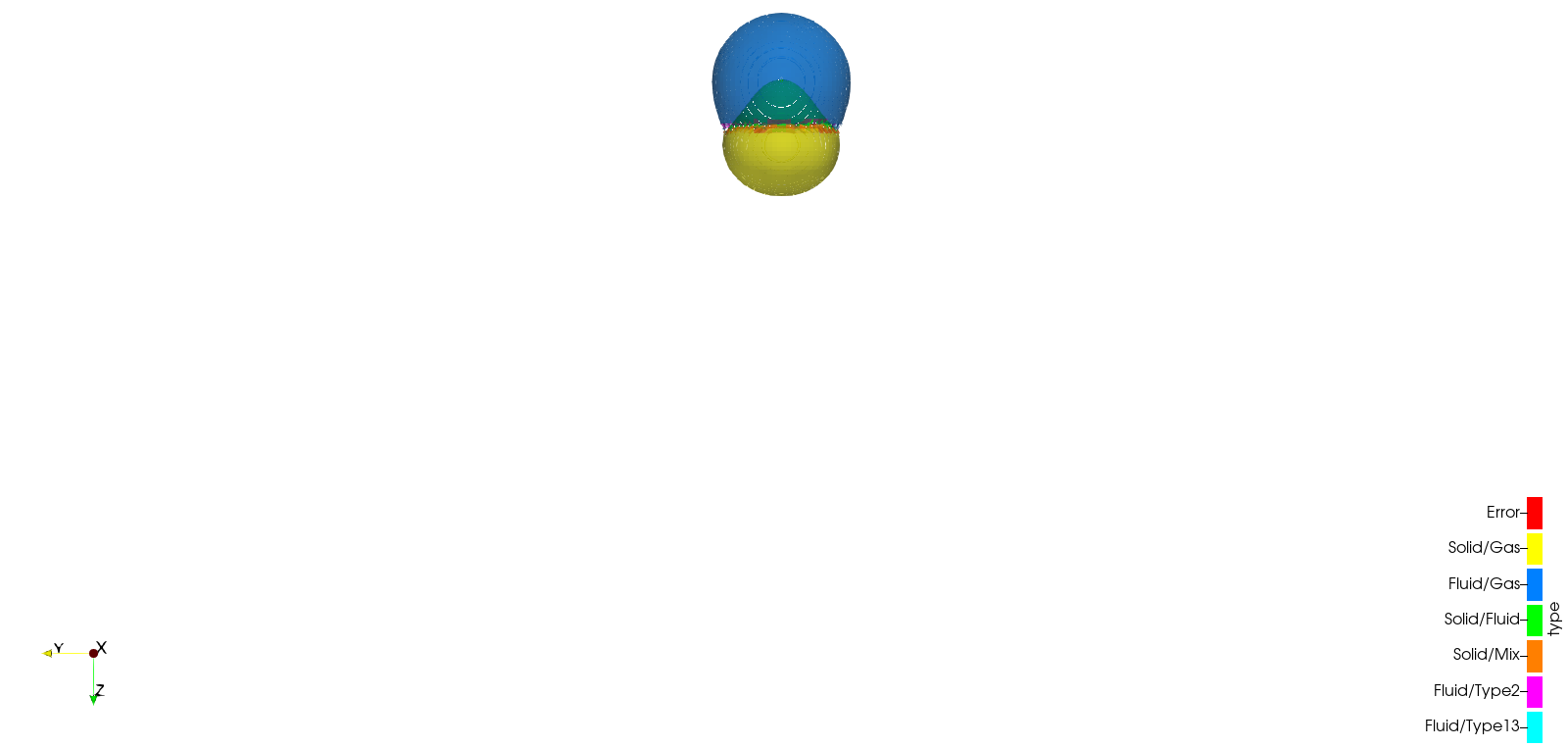}
\caption{$t=75~\mathrm{\mu s}$}
\end{subfigure}
\hfill
\begin{subfigure}[t]{0.19\linewidth}
\centering
\includegraphics[scale=0.4, trim = 25cm 16cm 25cm 1cm,clip]{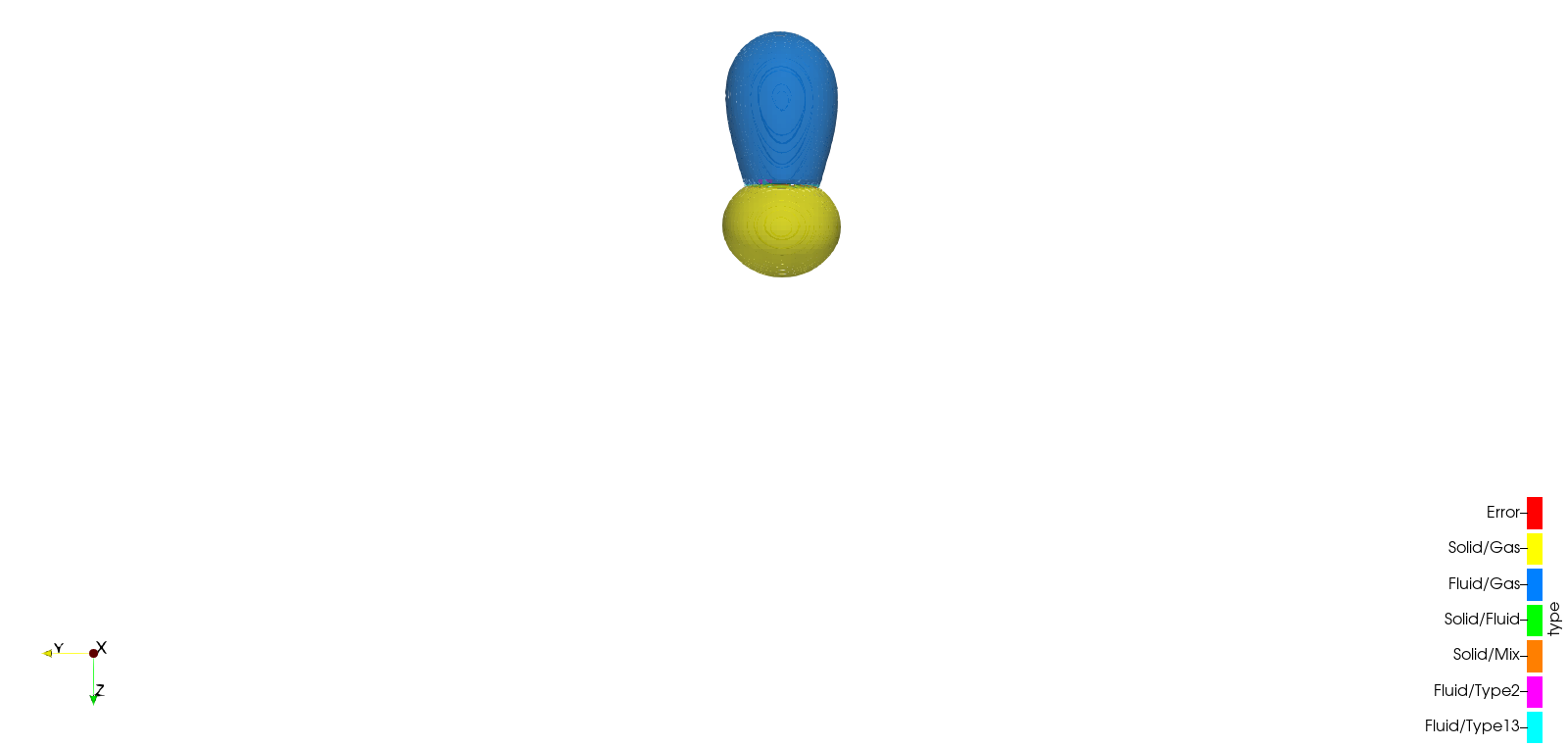}
\caption{$t=200~\mathrm{\mu s}$}
\label{Fig:HydrophobicPicC}
\end{subfigure}
\hfill
\begin{subfigure}[t]{0.19\linewidth}
\centering
\includegraphics[scale=0.4, trim = 25cm 11cm 25cm 5cm,clip]{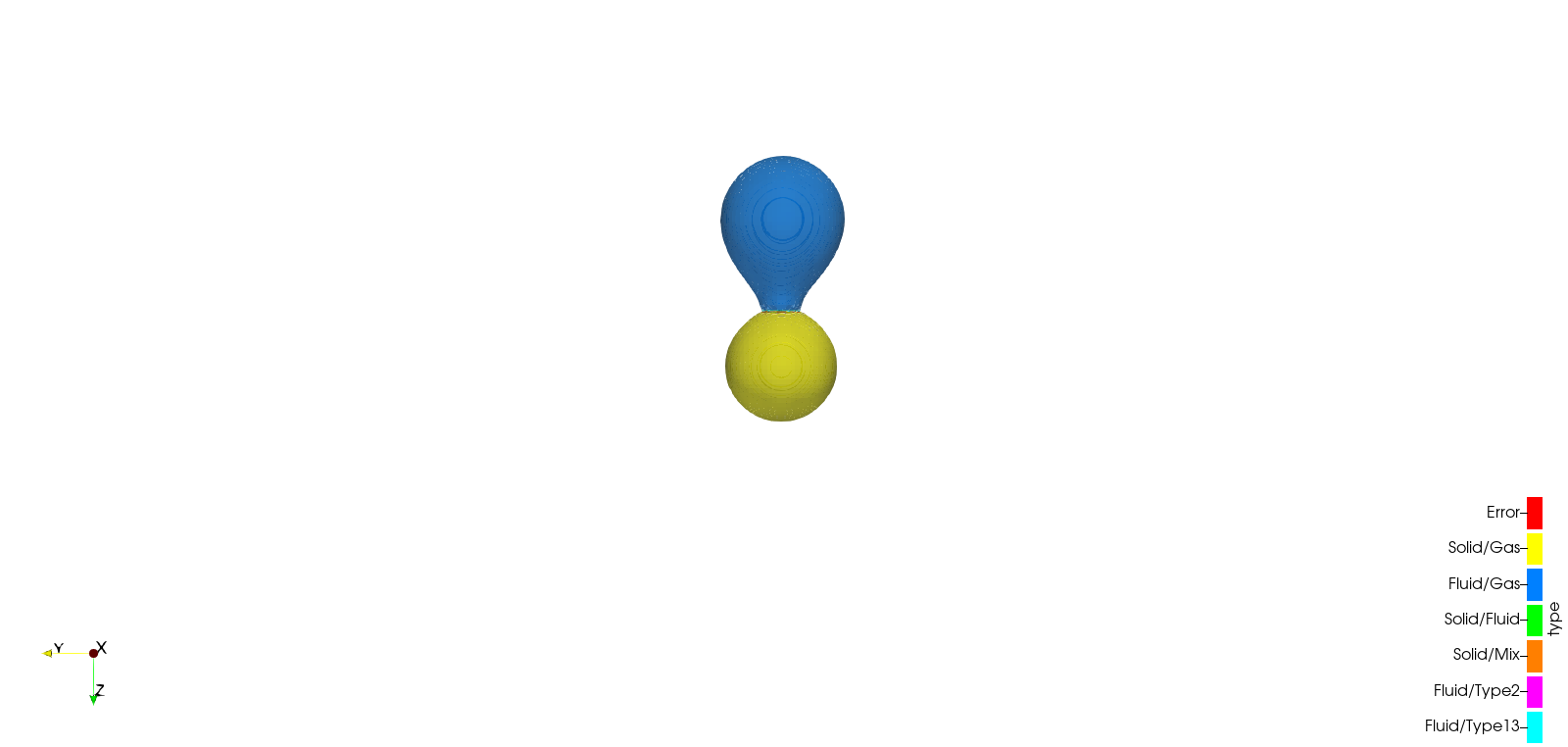}
\caption{$t=376~\mathrm{\mu s}$}
\label{Fig:HydrophobicPicD}
\end{subfigure}
\hfill
\begin{subfigure}[t]{0.19\linewidth}
\centering
\includegraphics[scale=0.4, trim = 25cm -1cm 25cm 20cm,clip]{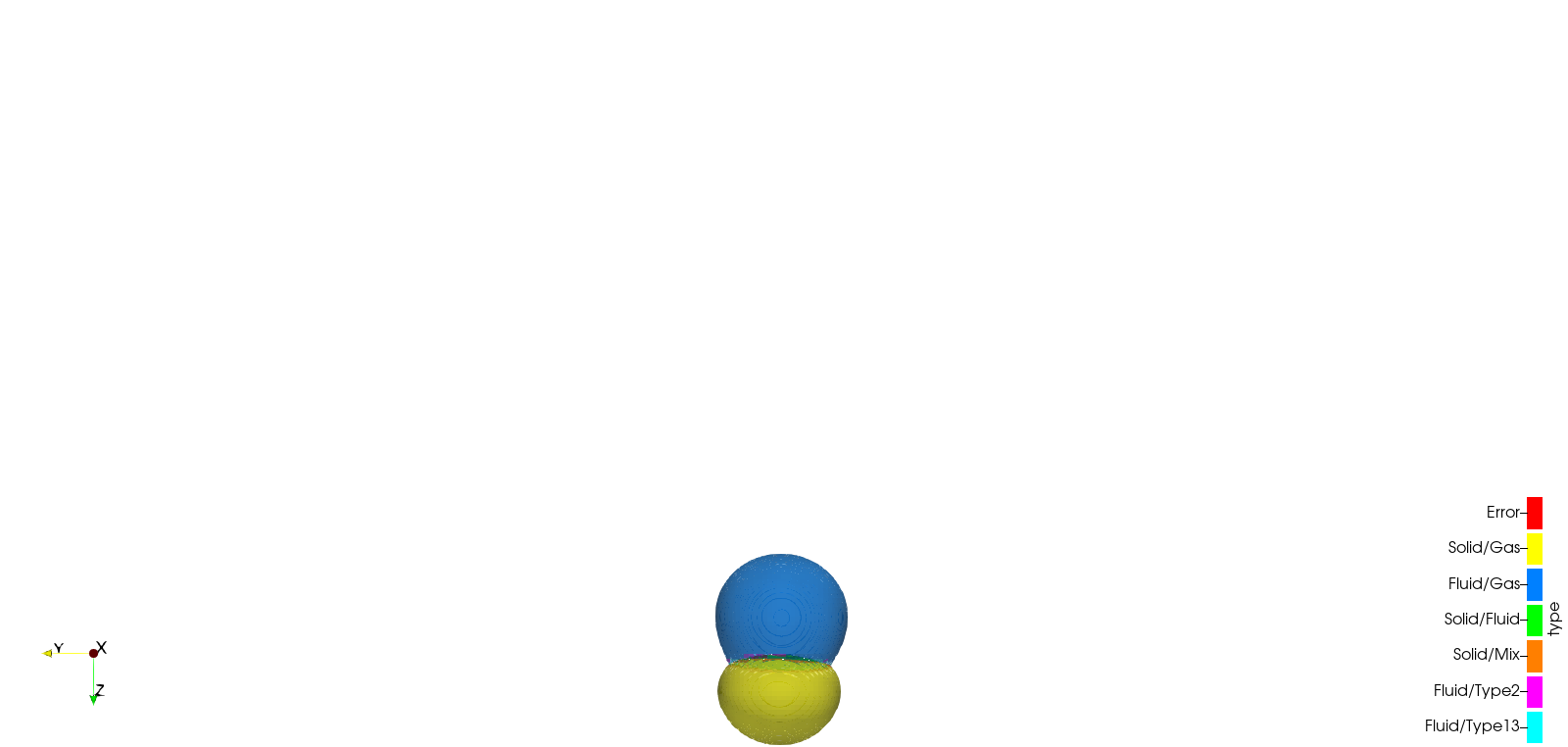}
\caption{$t=725~\mathrm{\mu s}$}
\end{subfigure}
\hfill\null
\centering
\begin{minipage}[t]{\linewidth}
\centering
\vspace{0.5cm}
\begin{tabular}{c c|c c}
\hline
$\sigma_{l_1-g}$ & $30.0~\mathrm{m N/m}$ & $\sigma_{l_2-g}$ & $40.0~\mathrm{mN/m}$\\$\sigma_{l_1-l_2} $ & $50.0~\mathrm{mN/m}$ & $\theta$ & $120^\circ$ \\
\hline
\end{tabular}
\end{minipage}
\caption{Partial wetting at a high contact angle: The initialisation is at rest. The contact line is retracting due to the interfacial forces until it reaches an equilibrium at the given contact angle. The interfaces are shown transparent and the colouring is analogous to the reconstruction test cases (liquid~1-gas yellow; liquid~1-liquid~2 green and orange in three-component cells; liquid~2-gas blue, magenta with liquid~1 inside the stencil  and cyan in three-component cells). Table~\ref{Tab:GeometryProperties1} show the parameters which are equal to the fully wetting test case's initialisation, the interfacial tensions which differ are given here.} \label{Fig:HydrophobicEllipsoids}
\end{figure}
\begin{figure}[tb]
\centering
\newcommand\filepath{.}
\begin{subfigure}[t]{0.19\linewidth}
\includegraphics[scale=2.2, trim = 175 130 175 0,clip]{\filepath/FullyWettingEllipsoids_00.png}
\caption{$t=0~\mathrm{ms}$}
\end{subfigure}
\begin{subfigure}[t]{0.19\linewidth}
\includegraphics[scale=2.2, trim = 175 125 175 5,clip]{\filepath/FullyWettingEllipsoids_05.png}
\caption{$t=125~\mathrm{\mu s}$}
\end{subfigure}
\begin{subfigure}[t]{0.19\linewidth}
\includegraphics[scale=2.2, trim = 175 120 175 10,clip]{\filepath/FullyWettingEllipsoids_10.png}
\caption{$t=250~\mathrm{\mu s}$}
\end{subfigure}
\begin{subfigure}[t]{0.19\linewidth}
\includegraphics[scale=2.2, trim = 175 95 175 45,clip]{\filepath/FullyWettingEllipsoids_20.png}
\caption{$t= 500~\mathrm{\mu s}$}
\label{Fig:HydrophilicEllipsoidsPicD}
\end{subfigure}
\begin{subfigure}[t]{0.19\linewidth}
\includegraphics[scale=2.2, trim = 175 5 175 130,clip]{\filepath/FullyWettingEllipsoids_43.png}
\caption{$t=1075~\mathrm{\mu s}$}
\label{Fig:HydrophilicEllipsoidsPicE}
\end{subfigure}
\centering
\begin{subfigure}[t]{\linewidth}
\centering
\vspace{0.5cm}
\begin{tabular}{c c|c c}
\hline
$\sigma_{l_1-g}$ & $68.6~\mathrm{mN/m}$ &
$\sigma_{l_2-g}$ & $19.5~\mathrm{mN/m}$\\
$\sigma_{l_1-l_2} $ & $34.3~\mathrm{mN/m}$ & $\theta$ & $\left(10^{-12}\right)^\circ$ \\
\hline
\end{tabular}
\end{subfigure}
\caption{Fully wetting test case: The contact angle used in the film stabilisation PLIC algorithm is $\left(10^{-12}\right)^\circ$ to account for the existence of a contact line also for fully wetting behaviour. The depiction is transparent to show the inner liquid's interface. The same color code as for the reconstruction test case is utilised (liquid~1-gas yellow; liquid~1-liquid~2 green and orange in three-component cells; liquid~2-gas blue, magenta with liquid~1 inside the stencil  and cyan in three-component cells). The film stabilisation for the reconstruction and the CSS model enables a thin film of liquid~2 spreading on liquid~1. Table~\ref{Tab:GeometryProperties1} show the parameters which are equal to the partially wetting test case's initialisation, while the interfacial tensions are given here.} \label{Fig:HydrophilicEllipsoids}
\end{figure}
\paragraph{Test Cases for Full and Partial Wetting} The wetting behaviour at equilibrium can be determined by the spreading parameter

\begin{equation}
S = \sigma_{l_1-g}-\sigma_{l_2-g}-\sigma_{l_1-l_2} \text{.}
\label{Eq:SpreadingParameter}
\end{equation}
When $S > 0$ the liquid~2 will fully wet liquid~1, if $S < 0$ only partial wetting occurs. Like indicated in Sec.~\ref{Sec:Introduction}, there is no extensive research data available on the wetting behaviour during the collision of immiscible droplets. Therefore, here two artificial test cases are used to prove that the surface forces are acting such that both behaviours, full encapsulation for fully wetting as well as a retraction of the contact line with a high contact angle $\theta = 120^\circ$ for partial wetting, are possible with the described methods. Both test cases are identically initialised. The same initialisation of the geometry with an initially moderate contact angle, density and viscosity, but different interfacial tensions, yields the two different outcomes. For ensuring a consistent choice of the static contact angle imposed and the interfacial forces in the simulation of partial wetting, the Young's Equation \cite{YOUNG1805} adapted from solid-wall interaction to immiscible liquids here as

\begin{equation}
\cos(\theta) = \frac{\sigma_{l_1-g}-\sigma_{l_1-l_2}}{\sigma_{l_2-g}}
\end{equation} 
is fulfilled. This ensures that the given static contact angle and the interfacial tensions are consistent at equilibrium. A fully wetting liquid combination is confirmed by a spreading parameter $S>0$, cf. Eq.~(\ref{Eq:SpreadingParameter}), and a contact angle $\theta=(10^{-12})^\circ$ is imposed then throughout the simulations. The dynamic contact line behaviour is omitted here, but imposing a varying dynamic contact angle from a correlation is possible in analogy to various three-phase studies, e.g. by G\"ohl~et~al.~\cite{GOEHL2018} or Patel~et~al.~\cite{PATEL2017}. Prominent examples of such correlations are e.g. presented by Kistler \cite{KISTLER1993} or Cox \cite{COX1986}. At the present, no dynamic contact line models for immiscible liquids in air are known to the authors. For fully wetting liquid combinations a static contact angle close to zero yields excellent results which is shown below in Sec.~\ref{Sec:ComparisonExperiment}, but a dynamic contact line modelling may be still of relevance for partially wetting liquids. Both test cases are initialised identically with the properties and geometry given in Tab.~\ref{Tab:GeometryProperties1}, while the interfacial forces are adapted for the different wetting behaviours as described above. Different outcomes can be observed for the two test cases for partially and fully wetting liquids. For the partially wetting liquid combination shown in Fig.~\ref{Fig:HydrophobicEllipsoids}, the repelling forces are reversed to attracting forces when the curvature exceeds the one imposed at the contact line by the surface tension combination, thus -geometrically- the imposed contact angle. This leads to the shape in Fig.~\ref{Fig:HydrophobicPicD}, where the liquid~2 is stretched close to the interface, as the surface forces are now working as attractive forces against the built up inertia. After this overshoot due to the built up inertia the equilibrium state given by the surface forces and the contact angle is reached eventually. For the fully wetting case in Fig.~\ref{Fig:HydrophilicEllipsoids}, the encapsulating forces are also pronounced, leading to a spreading of a thin film resulting in a fully encapsulated  compound droplet, cf. Fig.~\ref{Fig:HydrophilicEllipsoidsPicD}. After the full encapsulation,the inner droplet moves towards the centre of the outer droplet as the energetically most convenient position, cf. Fig~\ref{Fig:HydrophilicEllipsoidsPicE}. These two test cases show, that the topology capturing PLIC algorithm in combination with the CSS model modifications for immiscible liquids and thin films enable the simulation of fully and partially wetting immiscible liquids in contact.
\subsection{Interaction of Fully Wetting Immiscible Liquids}
\label{Sec:ComparisonExperiment}
\begin{table}[b]
\centering
\begin{tabular}{c c|c c|c c}
\hline
$\rho_{l_1} $ & $1126~\mathrm{kg/m^3}$ & $\sigma_{l_1-g}$ & $68.6~\mathrm{mN/m}$ &
 $\rho_{air}$ & $1.19~\mathrm{kg/m^3}$ \\
$\rho_{l_2} $ & $913.4~\mathrm{kg/m^3}$ & $\sigma_{l_2-g}$ & $19.5~\mathrm{mN/m}$&
 $\mu_{air}$ & $ 18.24 \cdot 10^{-6}~\mathrm{kg/(m~s)}$ \\
$\mu_{l_1} $  & $6.0\cdot 10^{-3}~\mathrm{kg/(m~s)}$ & $\sigma_{l_1-l_2} $ & $34.3~\mathrm{mN/m}$ &
 & \\
$\mu_{l_2} $ & $4.57\cdot 10^{-3}~\mathrm{kg/(m~s)}$ &  $\theta$ & $\left(10^{-12}\right)^\circ$ &
 & \\
\hline
\end{tabular}
\caption{Properties employed for the simulations of immiscible droplet collisions depicted in Fig.~\ref{Fig:CompareExperiments}. Liquid~1 is a 50\% Glycerol-Water solution, liquid~2 is Silicon Oil M5. The properties chosen are the same as in the experiments.}
\label{Tab:Properties}
\end{table}
\begin{figure}
\centering
\begin{minipage}[t]{\linewidth}
\includegraphics[scale=1, trim = 0cm 0.25cm 0cm 0cm, clip, page=1]{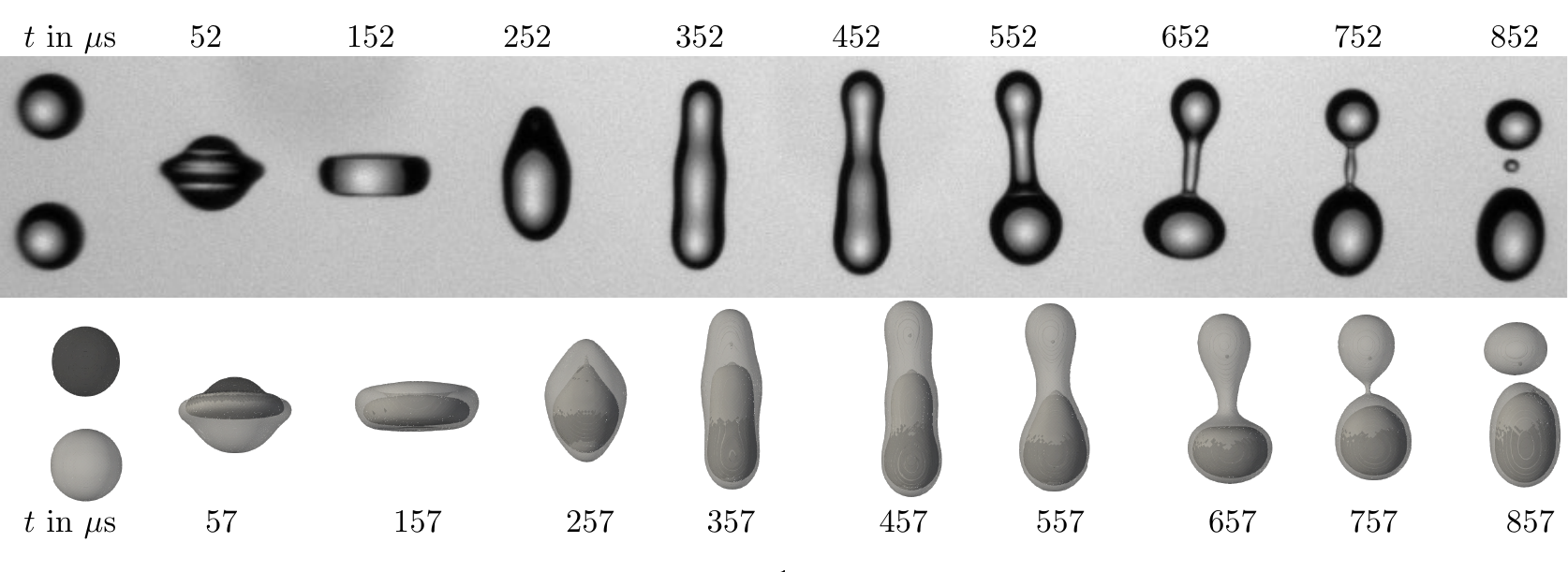}
\caption*{Crossing Separation: $d_{l_1}=188~\mu m$, $d_{l_2}=198~\mu m$, $X=0.0058$, $u_{rel}=3.47~m/s$; Domain: $1.2 \times 1.2 \times 0.8~\mathrm{mm}$ with a resolution of $192 \times 192 \times 128$ cells.}
\end{minipage}
\begin{minipage}[t]{\linewidth}
\includegraphics[scale=1, trim = 0cm 0.15cm 0cm 0.5cm, clip, page=2]{Planchette_G50SOM5_Vergleich.pdf}
\caption*{Coalescence: $d_{l_1}=188~\mu m$, $d_{l_2}=194~\mu m$, $X=0.027$, $u_{rel}=2.98~m/s$; Domain: $1.2 \times 1.2 \times 0.8~\mathrm{mm}$ with a resolution of $192 \times 192 \times 128$ cells.}
\label{Fig:CompareExperimentsCoalescence}
\end{minipage}
\begin{minipage}[t]{\linewidth}
\includegraphics[scale=1, trim = 0cm 0.25cm 0cm 0cm, clip, page=3]{Planchette_G50SOM5_Vergleich.pdf}
\caption*{Stretching Separation: $d_{l_1}=186~\mu m$, $d_{l_2}=194~\mu m$, $X=0.515$, $u_{rel}=1.97~m/s$; Domain: $1.2 \times 1.2 \times 0.8~\mathrm{mm}$ with a resolution of $192 \times 192 \times 128$ cells.}
\label{Fig:CompareExperimentsSeparation2}
\end{minipage}
\caption{Examples of the morphological comparison of simulation results of crossing separation, coalescence and stretching separation of Silicon Oil M5 (light) and a 50\% Glycerol-water solution (dark) with experimental results. Pictures of experiments provided by courtesy of C.~Planchette. The experimental method is described in detail in  \cite{PLANCHETTE2009,PLANCHETTE2010,PLANCHETTE2011,PLANCHETTE2012}. The properties of the liquids and air are shown in Tab.~\ref{Tab:Properties}.}
\label{Fig:CompareExperiments}
\end{figure}
In Sec. \ref{Sec:Introduction} it was described, that experimental data for the validation of simulations of immiscible binary droplet collisions is rare in literature. Chen and Chen \cite{CHEN2006} as well as Planchette~et.~al.~\cite{PLANCHETTE2009,PLANCHETTE2010,PLANCHETTE2011,PLANCHETTE2012} show experimental data, for fully wetting immiscible liquids. Therefore, a comparison in order to validate the simulation of droplet collisions of fully wetting immiscible liquids is possible. Planchette~et~al. report the interfacial tension for all three interfaces, Chen and Chen only the interfacial tensions towards the gaseous phase. Therefore, experiments by Planchette et al. are used here for a comparison to numerical simulations as there are no uncertainties arising from unknown liquid properties. \\ Figure~\ref{Fig:CompareExperiments} shows a comparison of collisions with different dimensionless impact parameter

\begin{equation}
X = \frac{b}{0.5\left(d_{l_1}+d_{l_2}\right)}\text{,}
\end{equation} 
thus the offset $b$ varies. The impact parameter as well as the relative velocity $u_{rel}$ affects the collision's outcomes: Crossing separation, stretching separation and coalescence. The fully wetting liquid combination Silicon Oil M5 (SOM5) and a 50\% Glycerol-Water solution (G50) were utilised for the experiments. The liquid's properties given in Planchette~et~al.~\cite{PLANCHETTE2011} and reproduced in Tab.~\ref{Tab:Properties} were imposed in the simulations accordingly. In the simulations, the exact diameters $d_{l_1}$ and $d_{l_2}$ as well as the relative velocities $u_{rel}$ measured in the experiments , provided by courtesy of C.~Planchette along with the pictures of the experimental results in the comparison, were initialised. In the simulations, gravitation was neglected to reduce the size of the computational domain required. It is believed, that especially during the first stages of collision the effect of the droplets falling is negligible. The comparison of the simulation results to the experimental data shows excellent agreement. The head-on regime boundary, the critical relative velocity $u_{rel,crit}$, between coalescence and crossing separation in the experiments is reported to be between $3.2~\mathrm{m/s}$ and $3.3~\mathrm{m/s}$ according to Planchette~et.~al.~\cite{PLANCHETTE2011,PLANCHETTE2012}, while in the simulation the regime boundary is slightly higher at $3.45~\mathrm{m/s}$. This is a prediction error of $4.5$ to $7.8~$\% compared to the minimum and maximum regime threshold from the experiments. The  critical impact parameter for off-centre collisions was found to be $X=0.495$ in the simulations, while in the experiment the threshold between coalescence and stretching separation is  at $X\approx0.5$ at a velocity of $u_{rel} = 1.97-1.98~m/s$. This  is a prediction error of $1$\% for this velocity. Additionally, the morphology comparison shows excellent agreement. The slight deviations in the shape in the middle of the collision's recording, cf. pictures at $260~\mu s$ and $360~\mu s$ in Fig.~\ref{Fig:CompareExperiments}, can be explained by the free fall experienced in the experiment which is not present in the simulations. Furthermore, the free fall seems to damp the induced rotation severely. Thus, it was necessary to adjust the orientation of the pictures from the simulations in $4-6^\circ$ steps up to $31^\circ$ in the last picture in this off-centre collision comparison. The effect of the free fall can be further investigated after a parallelisation of the implementation enables increased domain sizes. Throughout the collision, a contact angle $\theta=\left(10^{-12}\right)^\circ$ is imposed to account for the fully wetting behaviour and at the same time ensure, that the surface orientation at the contact line is always directed away from the bulk of liquid~2 to avoid numerical ``pearling-off" of small amounts of liquid~2 due to rounding errors in the interfaces orientation for parallel alignment. The simulations' results indicate, that a dynamic contact angle modelling is not necessary for this fully wetting liquid combination of SOM5 and G50. It should be noted that for other liquid combinations, especially for partially wetting liquids, this may not be the case. The slight deviation in the regime prediction may be due to the neglected gravity. At the late stages of the collision, the flow around the droplets may assist to drive the droplets further apart. It is also possible, that a further increase in the resolution will ensure, that the droplets connected through the filament are less influenced by the opposite interface, which is still inside the smoothing stencil as the filament connecting the two droplets becomes very short. Also, the very thin filaments, which develop in off-centre collisions, can be simulated more accurately with increased resolution. However, the present results show, that this details are not expected to change the prediction of the regimes much, as the prediction is already close to uncertainties present in experiments. \\
The present study revealed, that the number of smoothing steps leading to optimal results may be dependent on the resolution. With half the resolution, no smoothing at all was leading to the best comparison towards the experiments, while at the resolution presented here one smoothing step was sufficient. The smoothing might be optimal, when it is always performed in a zone of a certain constant width around the interface. The influence of the opposite interfaces and the dependence of the optimal smoothing on the grid resolution are to be studied after a parallelisation for supercomputing of the implementation in FS3D will enable a large increase in the resolution and domain size. 

\section{Summary and Outlook}\label{Sec:Conclusion}
New methods for fully three-dimensional numerical simulations of droplet interaction of immiscible li\-quids with different wetting behaviours were developed. An explicit algorithm based on different PLIC methods allows an efficient and interface reconstruction in three-component cells. It outperforms iterative methods regarding computational effort and is in terms of accuracy superior to explicit algorithms from literature due its applicability to arbitrary topologies. An extension of the two-phase CSS model to immiscible liquids was performed by introducing partial interfacial tensions, which is a method already known in literature. This method was further enhanced in this work with the introduction of a new film stabilisation. The interaction of droplets of immiscible liquids might result in a (partial) encapsulation of the liquid with the higher surface tension by the other one with a thin film. This situations with thin encapsulating films and moving contact lines can be simulated now in three dimensions without a grid refinement in the vicinity of the three-component cells.\\
Academic test cases proved the accurate reconstruction and advection of different topologies as well as the partial- and fully wetting behaviour which is  governed by the interfacial tensions only.\\
Quantitative validation data from experiments is available for the liquid combination Silicon Oil M5 and 50\% Glycerol-water solution which has a fully wetting behaviour. Excellent agreement was achieved in a comparison of the morphology and the regime transitions for head-on as well as off-centre collisions.\\
A parallelisation of the presented approach will allow for studies with higher resolutions and in larger domains, which are required e.g. to investigate the influence free fall on the collision's outcome. Hence, this approach provides the framework for the modelling and simulation of complex interactions of immiscible liquids.

\section*{Acknowledgements}
We kindly acknowledge the financial support by the Deutsche Forschungsgemeinschaft (DFG,
German Research Foundation) - Project SFB-TRR 75, Project number 84292822.\\
We would like to thank Prof. Dr. Dieter Bothe (Technical University of Darmstadt) for his valuable recommendations on the calculation of the surface forces which provided the basis for the successful modelling. Further thanks go to Prof. Dr. Ilia Roisman (Technical University of Darmstadt) for helpful suggestions on this topic.\\
We thank Ass.~Prof.~Dr.~Carole Planchette (Graz University of Technology) for contributing her expertise in many discussions on the physics of immiscible liquid interaction as well as providing high resolution pictures of experiments which made a validation possible.

\bibliography{mybibfile}

\end{document}